\documentclass[fleqn,usenatbib]{mnras}

\usepackage{newtxtext,newtxmath}
\usepackage{adjustbox}

\usepackage[utf8]{inputenc}
\usepackage[T1]{fontenc}
\usepackage{CJKutf8}

\DeclareRobustCommand{\VAN}[3]{#2}
\let\VANthebibliography\thebibliography
\def\thebibliography{\DeclareRobustCommand{\VAN}[3]{##3}\VANthebibliography}

\usepackage{graphicx}	
\usepackage{float}
\usepackage{amsmath}	
\usepackage[normalem]{ulem}
\usepackage[dvipsnames]{xcolor} 
\usepackage[normalem]{ulem} 

\usepackage{pdflscape}
\usepackage{array}
\usepackage{subfigure}

\newcommand{\ie}[0]{$\textnormal{i.e.}$}

\title[Chemical Diversity of the Metal-Poor MW]{The Chemical Diversity of the Metal-Poor Milky Way}

\author[N. Buckley et al.]{
Nicole Buckley,$^{1}$\thanks{E-mail: nicole.buckley0810@gmail.com}
Payel Das,$^{1}$
Paula Jofr\'e,$^{2}$
Robert M. Yates$^{3}$
and Keith Hawkins $^{4}$
\\
$^{1}$Department of Physics, University of Surrey, Stag Hill, Guildford, GU2 7XH, UK\\
$^{2}$Instituto de Estudios Astrofísicos, Facultad de Ingeniería y Ciencias, Universidad Diego Portales, Av. Ejercito 441, Santiago, Chile\\
$^{3}$Centre for Astrophysics Research, University of Hertfordshire, Hatfield, AL10 9AB, UK\\
$^{4}$Department of Astronomy, The University of Texas at Austin, 2515 Speedway Boulevard Austin, TX 78712, USA\\
}

\date{Accepted XXX. Received YYY; in original form ZZZ}

\pubyear{2023}

\begin{document}
\label{firstpage}
\pagerange{\pageref{firstpage}--\pageref{lastpage}}
\maketitle

\begin{abstract}
We present a detailed study of the chemical diversity of the metal-poor Milky Way (MW) using data from the GALAH DR3 survey. Considering $17$ chemical abundances relative to iron ([X/Fe]) for $9,923$ stars, we employ Principal Component Analysis (PCA) and Extreme Deconvolution (XD) to identify $10$ distinct stellar groups. This approach, free from chemical or dynamical cuts, reveals known populations, including the accreted halo, thick disc, thin disc, and in-situ halo. The thick disc is characterised by multiple substructures, suggesting it comprises stars formed in diverse environments. Our findings highlight the limited discriminatory power of magnesium in separating accreted and disc stars. Elements such as Ba, Al, Cu, and Sc are critical in distinguishing disc from accreted stars, while Ba, Y, Eu and Zn differentiate disc and accreted stars from the in-situ halo. This study demonstrates the potential power of combining a latent space representation of the data (PCA) with a clustering algorithm (XD) in Galactic archaeology, in providing new insights into the galaxy's assembly and evolutionary history.
\end{abstract}



\color{black}
\section{Introduction}

The Milky Way's (MW) stellar components have distinct star formation histories (SFHs) that manifest in unique chemical signatures. Chemical diversity, which measures the spread of elemental abundances, can indicate contributions from both nucleosynthetic processes and accreted systems. Variations in chemical diversity are more significant in the stellar halo than in the disc. This is because the halo has accumulated stars from various smaller galaxies, each with distinct SFHs, whereas the disc has experienced more thorough chemical homogenisation due to effective material mixing \citep{fengEarlyTurbulentMixing2014, reddyElementalAbundanceSurvey2006}.

The thick disc, first described by \citet{gilmoreNewLightFaint1983}, consists of prograde stars ($L_{z} > 0$) with scale heights between $530 \pm 32$ pc and $630 \pm 29$ pc above the Galactic plane \citep{kumarStudyGalacticStructure2021}. These stars are older than those in the thin disc ($\geq 10$ Gyr), more metal-poor ($-2.2 \leq$ [Fe/H] (dex)$ \leq -0.5$), and show higher enhancements of $\alpha$ elements relative to iron ([$\alpha$/Fe] $> 0.2$ dex) \citep{fuhrmannNearbyStarsGalactic, freemanNewGalaxySignatures2002, navarroThickThinKinematic2011, haydenCHEMICALCARTOGRAPHYAPOGEE2015}. The thick disc likely formed quickly, marked by an early burst of star formation in the MW's history. Its formation theories include: (i) secular evolution through heating and primordial collapse \citep{prantzosOriginGalacticThin2023}, (ii) a significant merger, e.g. Gaia Enceladus Sausage (GES), influencing the proto-disc before the thin disc's formation \citep{helmiMergerThatLed2018, belokurovCoformationDiscStellar2018a, bignoneGaiaEnceladusAnalog2019, ciucaChasingImpactGaia2023}, and (iii) direct accretion of stars from satellite galaxies \citep{reddyElementalAbundanceSurvey2006}.

The thin disc dominates the stellar material in spiral galaxies, with a high and constant star formation rate (SFR) due to abundant gas, leading to many young stars ($< 8$ Gyr \citep{yuBurstyOriginMilky2021}). It is the low-$\alpha$ ([$\alpha$/Fe] $< 0.2$ dex \citep{navarroThickThinKinematic2011}) region of the galactic disc and is more metal-rich due to later gas accretion. The thin disc extends to heights of $\approx 230 \pm 20$ pc to $330 \pm 11$ pc \citep{kumarStudyGalacticStructure2021}.

The MW stellar halo comprises two main parts: the inner and outer halo \citep{carolloTwoStellarComponents2007, nissenTwoDistinctHalo2010, beersCaseDualHalo2012}. Within the inner halo ($R_{\mathrm{gal}} < 15 \mathrm{kpc}$), there are high-$\alpha$ \citep{ishigakiChemicalAbundancesMilky2012} and low-$\alpha$ components \citep{nissenTwoDistinctHalo2010, hawkinsRelativeAgesArich2014,nissenTwoDistinctHalo2012, sheffieldIDENTIFYINGCONTRIBUTIONSSTELLAR2012, bensbyElementalAbundanceTrends2003, jackson-jonesGaiaESOSurvey2014}. The high-$\alpha$ component includes in-situ stars, GES, and the heated high-$\alpha$ disc \citep{zolotovDualOriginStellar2009, nissenTwoDistinctHalo2010, tisseraStellarHaloesMilkyWay2014, pillepichBUILDINGLATETYPESPIRAL2015}, formed from bursty star formation triggered by mergers \citep{fernandez-alvarAssemblyHistoryGalactic2019, ledinauskasReignitedStarFormation2018, liuChemicalHomogeneityAtomic2019, emamiTestingRelationshipBursty2021}. The low-$\alpha$ component is comprised of disrupted satellites with extended, less intense episodes of star formation. The inner halo is primarily influenced by GES, a significant dwarf galaxy accreted $\approx 8-11$ Gyr ago, marking the MW's last major minor merger \citep{belokurovBiggestSplash2020}.

Due to the availability of large stellar spectroscopic surveys, e.g. GALAH \citep{buderGALAHSurveyThird2021}, APOGEE \citep{majewskiApachePointObservatory2017}, SDSS/SEGUE \citep{yannySEGUESPECTROSCOPICSURVEY2009}, LAMOST \citep{cuiLargeSkyArea2012} and H3 \citep{conroyMappingStellarHalo2019}, many studies have used integrals of motion and chemical abundances to cluster and disentangle systems with different histories. Dynamical tagging is one method used, in particular using quasi-conserved quantities such as orbital actions ($J_{\rm z}$, $J_{\rm R}$ and $L_{\rm z}$) and energy ($E$). Although these quantities are not conserved over long timescales, the $E - L_{Z}$ space has been instrumental in identifying distinct accreted structures such as GES \citep{belokurovCoformationDiscStellar2018a, helmiMergerThatLed2018, helmiMappingSubstructureGalactic2002}, Sequoia \citep{myeongEvidenceTwoEarly2019}, and Kraken (a possible pre-disc in-situ population) \citep{kruijssenKrakenRevealsItself2020, forbesReverseEngineeringMilky2020}. 

Chemical tagging is the process of grouping stars together based on chemical abundance ratios \footnote{[X/Y] $= \log(N_{X}/N_{Y})_{*} - \log(N_{X}/N_{Y})_{\odot}$, is the logarithmic ratio between two abundances with respect to solar abundances, where $N_{X}$ represents the number of `X' atoms per unit volume.} \citep{freemanNewGalaxySignatures2002}. It assumes that stars born in the same population are chemically similar to each other but chemically distinct from other stars born at different times and/or locations. This allows us to use their preserved chemical signatures as snapshots of the galaxy's accretion history, facilitating a detailed reconstruction of the MWs formation through chemical tagging \citep{robertsonLambdaColdDarkMatter2005, fontChemicalAbundanceDistributions2006, bedellChemicalHomogeneitySunlike2018}. 
\citet{buderGALAHSurveyChemical2022} analyses up to $30$ element abundances, using them to successfully distinguish GES from the in-situ populations and find that the chemical signatures of Mg, Si, Na, Al, Mn, Fe, Ni, and Cu, significantly differ from in-situ MW stars. Through Gaussian mixture models applied to these chemical abundances, they isolate $1049$ stars associated with GES. 

In the literature, two-dimensional hyperplanes, such as the Tinsley-Wallerstein hyperplane which features [$\alpha$/Fe] and [Fe/H] \citep{tinsleyStellarLifetimesAbundance1979} have been used. The early-type dwarf galaxies formed stars early and have predominantly old stellar populations. The later-types with ongoing star formation have shallower [$\alpha$/Fe] - [Fe/H] slopes, which separates it from the steeper gradients present for MW stars. This is due to a difference in SFR, which is a result of the relative size difference between the smaller satellite systems and the MW \citep{dasAgesKinematicsChemically2020}. 

Another informative two-dimensional plane is [Mg/Mn] - [Al/Fe], shown in \citet{dasAgesKinematicsChemically2020} to effectively distinguish between accreted populations and the metal-poor disc \citep{hawkinsUsingChemicalTagging2015, mackerethOriginAccretedStellar2019, price-jonesStrongChemicalTagging2020}. Mg and Mn are relatively pristine tracers of core-collapse supernovae (CCSNe) and type 1a supernovae (SNe1a), respectively \citep{matteucciIntroductionGalacticChemical2016}. Similarly to [$\alpha$/Fe], the contrasting timescales of these events make [Mg/Mn] an invaluable tool for decoding the SFHs of systems. As for [Al/Fe], accreted populations, such as GES \citep{hawkinsUsingChemicalTagging2015} exhibit sub-solar values of this ratio due to the metallicity dependence of the light odd-Z element production \citep{tingPrincipalComponentAnalysis2012, kobayashiGalacticChemicalEvolution2006}.

Exploring other nucleosynthesis channels, \citet{maneaChemicalDoppelgangersGALAH2023} emphasised the role of neutron-capture elements in distinguishing between disc stars, significantly reducing the number of chemical doppelgangers. The ratio of s-process to r-process elements, especially [Ba/Eu], is important here. Analysing [Ba/Eu] ratios within these galaxies reveals sub-solar values at low metallicities ([Fe/H] $< -1.7$ dex), followed by a marked increase as more massive stars dominate ISM enrichment \citep{lanfranchiEvolutionBariumEuropium2006}. The difference between light s-process (ls) elements, such as yttrium and lanthanum, and heavy s-process (hs) elements, such as Ba is also crucial. These elements are primarily synthesised during the AGB phase of stellar evolution, with the [hs/ls] ratio serving as a gauge for neutron source efficiency and s-process output. Lower metallicity in AGB stars typically results in a higher [hs/ls] ratio, indicating stronger neutron-capture activity and the significant influence of the 13C pocket \citep{kappelerSprocessNucleosynthesisnuclearPhysics1989, bussoNucleosynthesisAsymptoticGiant1999}. Due to the obscurity surrounding the sites and yields of neutron-capture elements, there has been little analysis of these heavier elements in abundance space. 

While two-dimensional abundance hyperplanes effectively highlight enrichment differences between the halo and disc, the boundary between the inner halo and metal-poor disc requires more information to break down the degeneracy- implying that this region is more chemically complex and has a greater chemical diversity.
To uncover the extent of the chemical diversity in the MW, techniques have been applied to reveal the number of dimensions in chemical space (hereafter referred to as $\mathcal{C}$-space, drawing from \citealt{freemanNewGalaxySignatures2002}), and the abundance ratios which are most effective at breaking the chemical degeneracy. Having a high-dimensional $\mathcal{C}$-space is heavily recommended in \citet{andrewsInflowOutflowYields2017} for the next generation of multi-element stellar abundance surveys.

The choice of appropriate abundances is key when trying to recover signatures of accretion events. In \citet{tingHowManyElements2022}, they determine that for an observational uncertainty of $\approx 0.01 - 0.015$ dex, $5 - 7$ abundances must be conditioned on. This is shown in \citet{andersDissectingStellarChemical2018}, who used t-SNE (t-distributed Stochastic Neighbor Embedding), a dimensionality-reduction technique, to dissect $\mathcal{C}$-space revealing insights into the chemical diversity of the disc and suggesting that it is formed from various progenitor systems and is chemically complex. In this paper, we investigate this chemical diversity and assess the chemical degeneracy between the inner halo and the metal-poor MW (defined by $\mathrm{[Fe/H]} \leq -0.5$ dex) in the GALAH survey. The focus on the metal-poor MW is due to the high level of overlap in in-situ disc, halo and accreted populations when viewed in two-dimensional abundance planes and to also find known without the use of dynamical or chemical selection cuts- other than a metallicity cut. We aim to reveal nucleosynthesis contributions in the metal-poor MW and identify reproducible chemical hyperplanes for clustering chemically similar groups. After applying XD to the PCA-transformed $\mathcal{C}$-space, we distinguish these groups by their chemodynamical properties and compare them with known structures. By examining the metal-poor MW's chemical diversity, we seek to determine: (i) the most effective chemical abundances for resolving chemical degeneracy in the disc and disc-halo region; (ii) the effectiveness of unsupervised clustering algorithms in identifying chemically coherent groups in the transformed $\mathcal{C}$-space; (iii) insights into the formation histories of the thick disc and stellar halo.

In Section \ref{sec:data} we discuss the selection of the $17$ element enhancements ([X/Fe]) from GALAH DR3, and the sample cuts. In Section \ref{sec:methods} we utilise PCA to transform a high-dimensional ($17$) $\mathcal{C}$-space to a lower one ($9$), on which Extreme Deconvolution (XD) is applied to find chemically coherent groups. In Section \ref{sec:results}, we evaluate the contributions of the $17$ [X/Fe] ratios to the variance of the $\mathcal{C}$-space, which is examined in Section \ref{sec:discussion}. 

\section{Data} \label{sec:data}

We first present the GALAH+ (GALactic Archaeology with HERMES) DR3 \citep{buderGALAHSurveyThird2021} data and then discuss the chemical elements that will be included to break down degeneracy in $\mathcal{C}$-space. The GALAH survey aims to explore the chemical and kinematic histories of the MW. Using the HERMES spectrograph at the Australian Astronomical Observatory's $3.9$m Anglo-Australian Telescope, GALAH is a high resolution ($R \approx 28,000$), optical survey. One of its core strengths is its large amount of data and determination of chemical abundances for up to $30$ different elements for five different nucleosynthesis channels. Their abundance determination workflow uses Spectroscopy Made Easy (SME), which is a spectral synthesis fitting code \citep{piskunovSpectroscopyMadeEasy2017}.

In this work, we select $17$ [X/Fe] ratios (X = Mg, Si, Ca, Ti, Ba, La, Y, Eu, Mn, Zn, Co, Cr, Cu, Ni, Al, K and Sc), which were taken from an initial set of $30$ provided by GALAH. These abundance ratios have been chosen for their reliability and ability to capture key nucleosynthesis signatures in metal-poor stars. Abundance ratio measurements that are not measured in the imposed quality cuts are not included. Some elements were deliberately not included in this analysis: O and Na (both are light elements that are affected by internal mixing and depletion \citep{cohenChemicalAbundanceInhomogeneities2004}). Additionally [Fe/H], which is the iron abundance, was not included as [Fe/H] is highly correlated with the abundance ratios [X/Fe] that are considered in this work. We briefly summarise the elements included here below.

\begin{itemize}

    \item $\mathrm{[Mg,Si,Ca,Ti/Fe]}$ - The [$\alpha$/Fe] ratio is widely used to differentiate between CCSNe, which enrich the ISM with $\alpha$-elements on short timescales, and SNe1a, which contribute iron-peak elements across longer timescales. The play-off between these nucleosynthesis processes leads to a characteristic `knee' in the [Fe/H]-[$\alpha$/Fe] plane, the location of which can indicate the mass of the system, as more massive galaxies retain and accumulate more metals before SNe1a begin to dominate \citep{hendricksMETALPOORKNEEFORNAX2014, deboerEpisodicStarFormation2014}. Previous works highlight the chemical dichotomy of [$\alpha$/Fe] in both the disc and halo of the MW, acting as key indicators of SFH and the timing of chemical enrichment \citep{tinsleyStellarLifetimesAbundance1979}. 
    
    \item $\mathrm{[Ba/Fe]}$ - As a hs-element, Ba features super-solar values in systems with extended SFHs, indicative of enhancement due to AGB nucleosynthesis around $10$ Myrs after star formation \citep{delosreyesSimultaneousConstraintsStar2022}. At low metallicities, Ba is synthesized via the r-process, contributing to the chemical diversity observed in ancient stellar populations \citep{bussoNucleosynthesisAsymptoticGiant1999, frebelNucleiCosmosTracing2018}.

    \item $\mathrm{[La,Y/Fe]}$ - Observations suggest that ls elements exhibit more scatter than their heavy counterparts and are more overabundant at low metallicities \citep{burrisNeutronCaptureElementsEarly2000, aokiSpectroscopicStudiesVery2005}. This scatter could be indicative of varied nucleosynthetic sites and processes contributing to their abundances, reflecting different chemical evolution paths.

    \item $\mathrm{[Eu/Fe]}$ - Eu is predominantly produced by the r-process and serves as an excellent tracer of neutron-capture processes. The sources of the r-process are currently thought to be CCSNe, which contributes to the elevated Eu-enhancement observed at low-metallicities, as well as neutron star mergers (NSMs). NSMs occur on similar delay timescales to SNe1a \citep{wanajoNeutronStarMergers2021}, which leads to an overall increase of Eu abundance in time. Although studies indicate that NSMs are rare events, the observed r-process abundance trends in dwarf galaxies can predominantly result from stochastic effects \citep{beniaminiPROCESSPRODUCTIONSITES2016, jiRprocessEnrichmentSingle2016, carrilloDetailedChemicalAbundance2022}. This pattern is evidenced in dwarf galaxies from the Local Group such as Carina and Draco, which exhibit notable Eu enrichment ([Eu/Fe] $\approx 0.5$ dex) as observed by \citet{lanfranchiEvolutionBariumEuropium2006}.

    \item $\mathrm{[Mn/Fe]}$ - Mn is a more pristine tracer of SNe Ia compared to iron, which has other formation channels. Studies have shown that [Mn/Fe] is underabundant in low-metallicity stars and increases towards solar values at higher metallicities \citep{feltzingManganeseTrendsSample2007, northManganeseDwarfSpheroidal2012}.

    \item $\mathrm{[Zn,Co/Fe]}$ - Zn and Co are interesting for their roles in hypernovae (HNe) nucleosynthesis, providing insights into the high-energy end of stellar end-of-life processes \citep{kobayashiOriginElementsCarbon2020}. High [Zn/Fe] ratios in very metal-poor stars suggest CCSNe as a major site for Zn production but the required quantities are more consistent with energetic HNe. In later phases of galactic evolution, the major Zn-production site likely switches from HNe to SNe1a. This dual contribution led to the solar isotope composition of Zn and solar [Zn/Fe] values in the disc \citep{tsujimotoEarlyChemicalEvolution2018}. Low [Zn/Fe] can imply formation in an external environment with a low star formation efficiency and where the gas has been poorly enriched by massive stars \citep{minelliNewSetChisels2021}.

    \item $\mathrm{[Cr,Cu,Ni/Fe]}$ - Cr is primarily produced in SNe1a and CCSNe, with its abundance reflecting the integrated yield of these events over time. Cu, on the other hand, has a more complex nucleosynthetic origin, involving both SNe1a and CCSNe, but also significant contributions from AGB stars. Ni is predominantly synthesised in SNe1a, making its abundance a valuable tracer for these events.

    \item $\mathrm{[Al,K,Sc/Fe]}$ - Al and K are included for their potential to reveal unique nucleosynthetic signatures from massive stars and AGB stars. The in-situ and accreted halo appear to exhibit different values of [Al/Fe], with metal-poor populations having a lower efficiency of Al production due to inefficient C, N and O in the ISM (see \citealt{kobayashiGalacticChemicalEvolution2006} for further details). 
    
\end{itemize}

We select stars with a sufficient spectrum signal-to-noise ratio (S/N), with no identified problems with stellar parameter determination, no identified problems with abundance determination of any of the elements we consider, and with a low uncertainty in abundance ratios by imposing the following cuts to the GALAH data (as recommended in \citealt{buderGALAHSurveyThird2021}):

\begin{itemize}

    \item \texttt{snr$\_$c3$\_$iraf} $> 30$
    
    \item \texttt{flag$\_$sp} $== 0$
    
    \item \texttt{flag$\_$Fe$\_$H} $== 0$
    
    \item \texttt{flag$\_$X$\_$Fe} $== 0$
    
    \item \texttt{e$\_$X$\_$Fe} $< 0.2$
    
\end{itemize}

Without these cuts, spurious relations can appear and distort the abundance relations by large residual correlations \citep{tingHowManyElements2022, jofreAccuracyPrecisionIndustrial2019a}. We only select red giant branch (RGB) stars ($\log{g} < 3$ and $\log{T_{\mathrm{eff}}} < 3.73$), due to their high intrinsic luminosity which allows us to probe more distant stars. By conditioning on the stellar parameters $T_{\mathrm{eff}}$ and $\log{g}$, we also ensure that our sample is not impacted by the different measurement systematics between spectral types \citep{jofreAccuracyPrecisionIndustrial2019a, demijollaDisentangledRepresentationLearning2021}. A final additional cut of [Fe/H] $\leq -0.5$ dex has been made to focus on metal-poor stars. After making these cuts, our sample size is reduced from $678,423$ to $9,923$ stars.

\begin{figure}
 \centering
    \includegraphics[width=\linewidth]{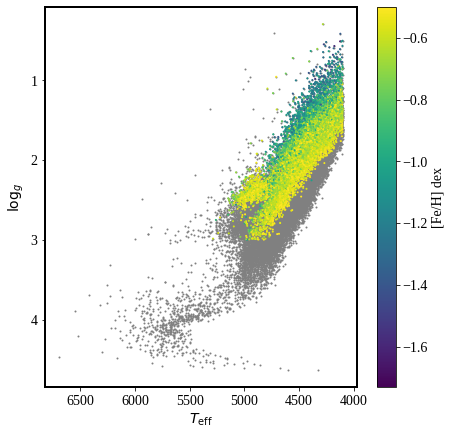}
 \caption{The logarithmic surface gravity and effective temperature of GALAH stars. The grey points represent the full sample (without the cuts discussed above), while overlaid points are of our working sample coloured by their metallicity. This limits our sample to RGB stars, which are selected due to their high luminosity.}
 \label{fig:HR_Diagram}.
\end{figure}

\section{Methods} \label{sec:methods}

In the following section, we describe the details of using PCA to find a compact latent space transformation of the high-dimensional $\mathcal{C}$-space. The transformation allows us to explore the importance of each element in the chemical diversity of the dataset and helps the XD clustering method by reducing the dimensionality of the dataset \citep{bovyExtremeDeconvolutionInferring2011, mukherjeeCapturingDenoisingEffect2024}.

\subsection{Principal Component Analysis (PCA)}\label{sec:PCA}

Principal Component Analysis (PCA) is a dimensionality reduction technique that forms linear combinations of the original variables, ordered by their information content (i.e. contribution to the total variance). By removing those principal components (PCs) with the least information, the data can be simplified in an optimal way, making it easier to visualise and allowing clustering algorithms to find groups more cleanly \citep{jolliffePrincipalComponentAnalysis2016}. 

We refer to our analysis as Principal Component Abundance Analysis (PCAA, coined by \citet{andrewsPrincipalComponentAbundance2012}). Previous studies have shown PCA's potential to reveal latent correlations between chemical elements, providing insights into nucleosynthetic pathways and the galaxy's chemical evolution \citep{tingPrincipalComponentAnalysis2012, andrewsPrincipalComponentAbundance2012, tingPayneSelfconsistentInitio2019}. This process yields eigenvectors (PCs) and eigenvalues representing the variance captured by each component. A larger eigenvalue indicates that its corresponding PC captures more variance, holding more `information' about the original data. By projecting onto a subset of the most significant eigenvectors, the data set can be dimensionally reduced. 

PCA's linearity, unlike t-SNE, necessitates that our data reflect this property. The decision to work in [X/Fe] space is based on two assumptions: (1) [X/Fe] correlations are linear due to their logarithmic nature; and (2) [X/Fe] errors are Gaussian distributed. [X/H] ratios are highly correlated, offering less distinctiveness in the $\mathcal{C}$-space. \citet{tingPrincipalComponentAnalysis2012} analysed the dimensionality of $\mathcal{C}$-space in various metallicity bands, revealing that neutron-capture abundances significantly contribute to observed variances in low-metallicity ranges ($-3.5 < \mathrm{[Fe/H]} < -1.5$ dex). They demonstrated that the dimensionality of $\mathcal{C}$-space, defined by the number of PCs required to encapsulate intrinsic variations, is accurately represented when the cumulative variance of the first $k$ PCs is around $85\%$. This threshold is influenced by sample size, observed abundances, and their uncertainties. Unlike two-dimensional chemical hyperplanes, PCA probes how different elements evolve together, revealing the chemical diversity of the MW.

\subsection{Extreme Deconvolution (XD)} \label{sec:XD}

In this study we employ XD, a clustering algorithm, designed to recover an underlying multi-Gaussian distribution, even in the presence of noise. We start with the assumption that the $N$ observed data points are drawn from a mixture of $k$ multivariate Gaussian distributions with means $\mu_{k} \in \mathbb{R}^{d}$ and covariances $\Sigma_{k} \in \mathbb{R}^{d \times d}$, where $d$ is the number of dimensions. For each observed data point $x_{i}$, there is an associated observational error, which can be modelled as a Gaussian. The covariance matrix of this error is represented as $\chi_{i}$. The initial predictions on the fraction of the total $N$ points that belong to the $k^{\mathrm{th}}$ Gaussian component are given by $\pi_{k}$, such that $\sum^{K}_{k=1}\pi_{k} = 1$. The initial estimates for $\pi_{k}$, $\mu_{k}$ and $\Sigma_{k}$ are initialised using the \verb|sklearn.mixture.GaussianMixture| package, and are generated using K-Means. The XD algorithm employs the Expectation-Maximisation (EM) technique, which consists of two steps: the E-step, where the posterior probabilities $\omega_{k,i}$ are computed, and the M-step, where the Gaussian parameters are updated. 

In the E-step, the algorithm calculates the probability that data point $x_{i}$ belongs to the $k^{\mathrm{th}}$ Gaussian component using Bayes' theorem:

\begin{equation}\label{eqn:Bayes_theorem}
\omega_{k,i} = \frac{p(x_{i}|\mu_{k}, \Sigma_{k} + \chi_{i})}{\sum_{j=1}^{k} \pi_{j} p(x_{i}|\mu_{j}, \Sigma_{j} + \chi_{i})},
\end{equation}
where $p(x_{i}|\mu_{k}, \Sigma_{k} + \chi_{i})$ represents the probability density function of the Gaussian distribution with mean $\mu_{k}$ and covariance $\Sigma_{k}$ evaluated at $x_{i}$. In the M-step, the estimates for $\mu_{k}$ and $\Sigma_{k}$ are updated. The revised mean $\mu_{k}$ for all $k$ is:
\begin{equation}\label{eqn:weighted_mean}
\mu_{k} = \frac{1}{N_{k}} \sum_{i=1}^{N} \omega_{k,i} x_{i},
\end{equation}
with $N_{k}$ being the effective count of data points ascribed to the $k^{\mathrm{th}}$ Gaussian component. The updated covariance estimate for all $k$ is:
\begin{equation}\label{eqn:covariances}
\Sigma_{k} = \frac{1}{N_{k}} \sum_{i=1}^{N} \omega_{k,i} (x_{i} - \mu_{k})(x_{i} - \mu_{k})^T - \chi_{i}
\end{equation}
This formula accounts for the observational error, refining the true distribution's covariance estimate. Lastly, the prior probability $\pi_{k}$ is adjusted:
\begin{equation}\label{eqn:predictions}
\pi_{k} = \frac{N_{k}}{N}
\end{equation}

The XD algorithm iterates between the E-step and M-step until convergence is reached. XD's robustness against noise and its ability to discern intricate structures make it particularly valuable in astrophysics. This was show cased by \citet{bovyExtremeDeconvolutionInferring2011} and \citet{belokurovCoformationDiscStellar2018a}, to break down the complex velocity distributions of solar neighbourhood stars and the MW stellar halo, respectively. 

\subsection{Group Number (k) Estimation} \label{sec:group_number}

After employing XD, the next critical task is to determine the optimal number of Gaussian groups, hereafter defined as $k$. The Bayesian Information Criterion (BIC) emerges as a practical tool for this purpose \citep{fraleyModelBasedClusteringDiscriminant2002}. It provides a trade-off between the fit of the model to the data and the dimensionality (hereafter defined as $d$) of the model. The formula for BIC is given by:
\begin{equation}\label{eqn:BIC}
\mathrm{BIC} = -2\ln (L) + m \ln (N),
\end{equation}
where $L$ is the likelihood of the observed data given the model, $m$ is the number of free parameters in the model and as before, $N$ is the number of data points. $L$ is the product of the probabilities of each data point, given the model parameters and is given by:
\begin{equation}\label{eqn:likelihood}
L = \prod^{N}_{i=1} p(\mathbf{x}_{i}|\mathbf{\theta}),
\end{equation}
where $\mathbf{\theta}$ denotes the set of parameters of the Gaussian mixture model. The value of $m$ is given by:
\begin{equation}\label{eqn:num_free_parameters}
m = d \frac{d+3}{2} k.
\end{equation}

The BIC tends to favour models with fewer parameters (simpler models) for smaller datasets and models with more parameters (more complex models) for larger datasets. As a result of this, the BIC penalises overfitting, discouraging overly complex models. In our study, under-estimating the number of groups is more advantageous than over-estimating, as during subsequent dynamical analyses these merged groups could be teased apart based on their dynamical properties.

\begin{table}
    \centering
    \begin{tabular}{|c|c|c|c|} \hline 
         $d$-dimensions&  $k$-groups&  $X_{\mathrm{err}}$ &  $k$-predicted \\ \hline 
         3&  3&  0.0, 0.05, 0.10, 0.20&  3, 3, 3, 3 \\ \hline 
         3&  5&  0.0, 0.05, 0.10, 0.20&  5, 4, 5, 6 \\ \hline 
         3&  10&  0.0, 0.05, 0.10, 0.20&  5, 5, 5, 4 \\ \hline 
         5&  3&  0.0, 0.05, 0.10, 0.20&  3, 3, 3, 2 \\ \hline 
         5&  5&  0.0, 0.05, 0.10, 0.20&  6, 6, 6, 6 \\ \hline 
         5&  10&  0.0, 0.05, 0.10, 0.20&  11, 11, 11, 8 \\ \hline 
         7&  3&  0.0, 0.05, 0.10, 0.20&  3, 3, 3, 3 \\ \hline 
         7&  5&  0.0, 0.05, 0.10, 0.20&  8, 5, 5, 5 \\ \hline 
         7&  10&  0.0, 0.05, 0.10, 0.20&  11, 6, 9, 9 \\ \hline
    \end{tabular}
    \caption{The compiled results of the XD mock tests, investigating the impact of the dimensionality ($d$), number of groups ($k$) and the errors $X_{\mathrm{err}}$.}
    \label{tab:mock_table}
\end{table} 

In order to test the accuracy and limitations of XD, we generated a set of mock multivariate distributions, adjusting the number of groups ($k$), means ($\mu_{k}$), covariances ($\Sigma_{k}$), weights ($\pi_{k}$), number of dimensions ($d = 2$, $5$ and $7$) and magnitude of the errors ($X_{\mathrm{err}} = 0.0, 0.05, 0.10, 0.20$ dex). The summarised results are found in Table \ref{tab:mock_table}. These errors were chosen to span the range of errors that are encountered in the GALAH data. To keep these tests applicable to the ranges of our PCA-transformed data, we use the same sample size ($N = 9,923$), generating random samples with the following number of artificially included groups: $3$, $5$, $7$ and $10$. XD performs well in recovering the correct number of groups, $k$, but this performance is influenced by both the dimensionality, $d$, of the space and the magnitude of $k$. In a three-dimensional space with three groups ($d=3, k=3$), the number of groups predicted by the BIC, $k_{\mathrm{predicted}}$, is accurate. However, when the number of groups is increased to $10$ in the same three-dimensional space the prediction becomes less reliable, resulting in $k_{\mathrm{predicted}}=4$, when the errors are set to $X_{\mathrm{err}} = 0.2$ dex. As the dimensionality increases, the accuracy of the group recovery seems to improve, for example, in the case of $d = 7$ and $k = 10$, the BIC is able to predict a range of values close to the actual number, $k_{\mathrm{predicted}}=6$ to $11$. 

\begin{figure}
 \centering
    \includegraphics[width=\linewidth]{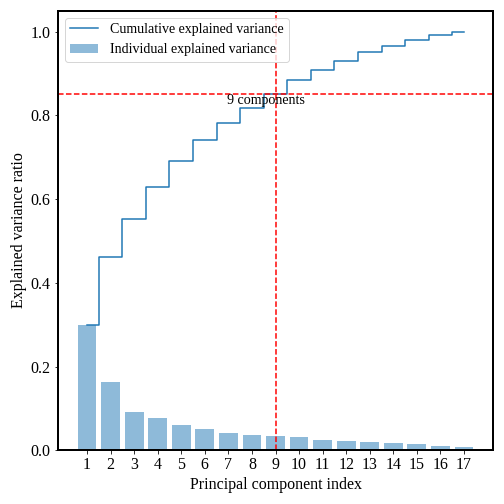}
 \caption{The explained variance of a PC is the eigenvalue of that PC. Applying a cut-off at $85\%$ cumulative explained variance shows that $9$ components are required to reliably capture the patterns present in the data. Thus a dimensionality reduction can be made from a $17$ to a $9$ dimensional space.}
 \label{fig:explained_variance}
\end{figure}

\begin{figure*}
  \centering
    \subfigure{
      \includegraphics[width=0.45\linewidth]{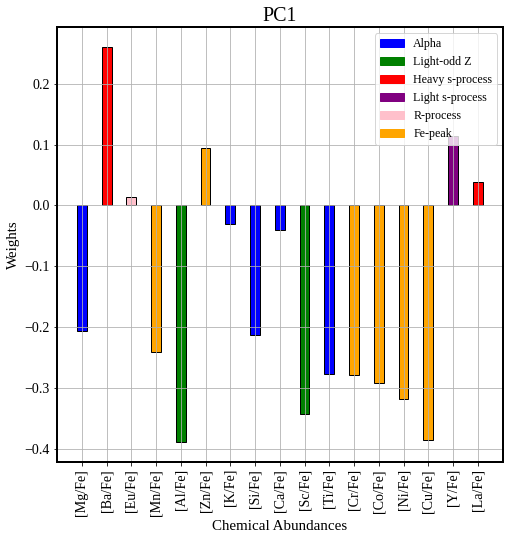}
    }
    \hspace{0.01\linewidth}
    \subfigure{
      \includegraphics[width=0.45\linewidth]{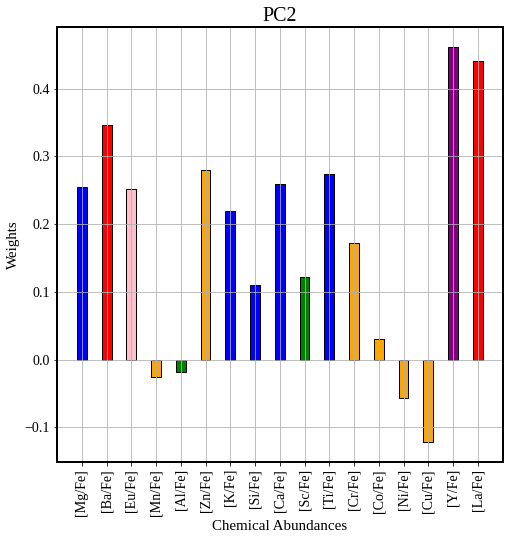}
    }
  \caption{The $17$ [X/Fe] abundances are coloured approximately by their nucleosynthetic families: $\alpha$ (blue), light-odd Z (green), heavy s-process (red), light s-process (purple), r-process (pink) and iron-peak (orange). The first PC (left plot) makes up $29\%$ of the total variance of the $\mathcal{C}$-space. The second PC (right plot) makes up $18\%$ of the total variance. The weights (positive and negative) show how each [X/Fe] chemical abundance ratio contributes to the PC.}
  \label{fig:PC1_PC2}
\end{figure*}

This suggests that higher-dimensional spaces provide better separation between groups, allowing for more accurate predictions by the BIC. In summary, when looking at various dimensions and group numbers, $k_{\mathrm{predicted}}$ does not consistently indicate a tendency toward overfitting or underfitting our mock tests. Nevertheless, in cases where $k=10$, which is the number used in this work as described in Section \ref{sec:cluster_analysis}, the predicted number of groups falls short across all dimensions, indicating a tendency towards underfitting. Consequently, we conclude that when applying this method to our data, the BIC score's `knee' (the point where adding more groups does not significantly improve the model) represents a lower limit of  the  true number of groups in our dataset, so is unlikely to subdivide the $\mathcal{C}$-space into too many groups.

\section{Results} \label{sec:results}

Here, we describe the results of applying PCA and our clustering analysis, followed by a presentation of the chemodynamical properties of the revealed groups- assigning known stellar components to the $10$ derived groups as follows: the thick disc (G1, G4, G6, G7 and G8), GES (G2), the in-situ halo (G3 and G5) and the thin disc (G9). G10 is interpreted as an outlier group.

\subsection{PCAA} \label{sec:PCAA}

PCA is applied to the $17$-dimensional $\mathcal{C}$-space, with the explained variance shown in Figure \ref{fig:explained_variance}. The explained variance is the fraction of the total variance in the original data set that is contained in each PC. Consistent with the approach of \citet{tingPrincipalComponentAnalysis2012}, we have implemented a cumulative explained variance cutoff of $85\%$, thereby excluding those PCs dominated by noise or errors. Notably, the cumulative explained variance increases gradually, requiring a substantial number of PCs to reach the $85\%$ threshold. This pattern could suggest, within the metallicity range of $-1.7 \leq \mathrm{[Fe/H]} \leq -0.5$, the presence of multiple independent nucleosynthetic processes that influence the observed chemical patterns, as well as different SFHs contributing to the enrichment of this region. These aspects are explored further in Section \ref{sec:discussion}. Implementing the $85\%$ cutoff reduces the $17$-dimensional space to a $9$-dimensional space, highlighted in Figure \ref{fig:explained_variance}. We note that changing the number of PCs we consider by a few does not significantly affect the number of groups that are identified by our subsequent clustering analysis in Section \ref{sec:cluster_analysis}.

\begin{figure}
 \centering
    \includegraphics[width=0.9\linewidth]{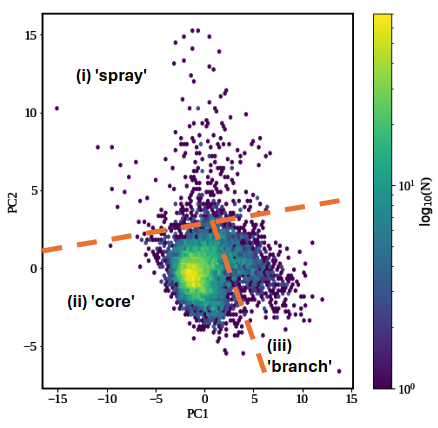}
 \caption{The `core' is the group centered around PC1 $\approx 0$ and PC2 $\approx -2$. The `branch' is the group centered around PC1 $\approx 5$ and PC2 $\approx 0$. The `spray' is the group centered around PC1 $\approx 0-1$ and PC2 $\approx 10$.}
 \label{fig:PCA_labelled}
\end{figure}

\begin{figure*}
    \centering
        \includegraphics[width=0.9\linewidth]{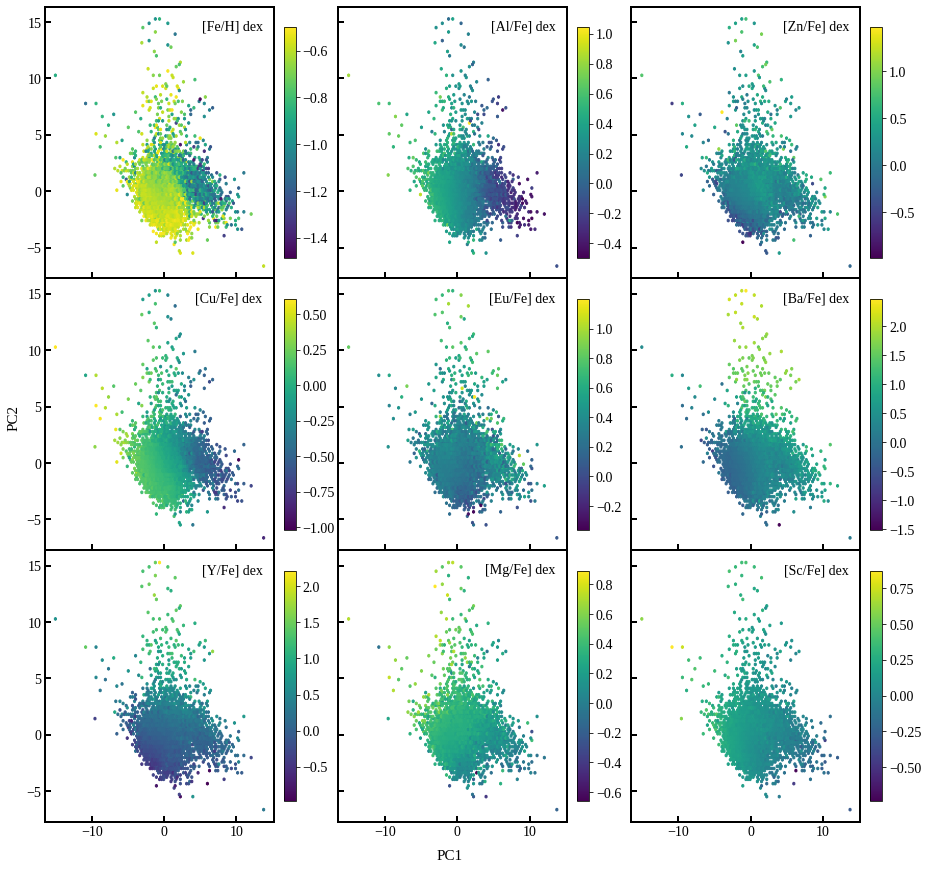}
        \caption{The PC1-PC2 hyperplane coloured by various abundances ratios ([Fe/H], [Al/Fe], [Zn/Fe], [Cu/Fe], [Eu/Fe], [Ba/Fe], [Y/Fe], [Mg/Fe] and [Sc/Fe]). The only abundance ratio here that was not included in the $\mathcal{C}$-space is [Fe/H]. We evaluate the chemical enrichment in three regions: the `core', the `branch' and the `spray'.}
    \label{fig:PC1_PC2_core_offshoot_spray}
\end{figure*}

To further breakdown the elemental contributions to the variance, Figure \ref{fig:PC1_PC2} shows the weights (\ie{} eigenvector coefficients for each element enhancement) for the first two PCs, which contain $50\%$ of the variance of the data. In PC1, [Al/Fe], [Cu/Fe] and [Ba/Fe] contribute the most to the variance (in terms of absolute magnitude). In PC2, there is a greater contribution from the ls-elements, [Y/Fe] and [La/Fe]. The lack of contribution, positive or negative, from elements like [Eu/Fe] and [K/Fe] in PC1, and [Mn/Fe], [Al/Fe], [Co/Fe], [Ni/Fe], and [Cu/Fe] in PC2, is a consequence of the non-linearity of some nucleosynthesis processes, e.g., Al, where two trends can cancel out. To see more clearly how these elements contribute to the spread of observed abundances, we show the PC1 - PC2 plane in Figure \ref{fig:PCA_labelled}. We use the following terms to define the three general regions that are identifiable by eye in this two-dimensional plane; the `core', the `branch' and the `spray'. We argue in the following that the `core' is made up of disc populations, the `branch' is made up from a single accreted population (GES) and the `spray' features super-solar s-process enhanced stars which are in-situ in origin. These regions are used for reference when analysing the distributions of each [X/Fe] abundance ratio in the PC1 - PC2 plane in Figure \ref{fig:PC1_PC2_core_offshoot_spray}. We can also identify that the positive PC1 direction varies between disc (`core') and accreted stars (`branch'), with hs-element [Ba/Fe] having a positive contribution and iron-peak and light odd-Z elements having a negative contribution which points towards metal-poor populations with extended SFHs. The positive PC2 direction varies between disc and in-situ populations (`spray'), pointing towards s-process and $\alpha$-enhancement. Our main findings are as follows:

\begin{itemize}

\item Core: The `core' is characterized by relatively metal-rich stars with [Fe/H] $\approx -0.6$ dex, noting that the sample is selected to have [Fe/H] $\leq -0.5$ dex. This region shows significant enhancement in [Al/Fe], indicating differences in SFHs compared to other regions- with the `core' experiencing recent/ongoing star formation compared with the accreted populations in the `branch'. The majority of the `core' shows enhanced [Cu/Fe], except for the lower region occupying PC1 $\approx -2$, PC2 $\approx -4$. The `core' has near-solar values of [Ba/Fe], suggesting it is representative of disc populations. There is a lower [Mg/Fe] value in the lower region of the `core' compared with the upper `core', with a difference of $\approx 0.3$ dex, which is greater than the average errors of [Mg/Fe] ($0.07$ dex). This could indicate the presence of thin disc stars. Additionally, some [Sc/Fe] enhancement is observed in the left region of the `core' (at PC1 $\approx -5$ and PC2 $\approx 0$).

\item Spray: The `spray' region displays a complex mixture of metallicities. It is notable for hs-enhancement in [Ba/Fe] and super-solar [Y/Fe] stars, indicating a distinct chemical signature from both the `core' and `branch'. PC2 shows a high positive weighting for s-process elements towards the `spray', suggesting significant contributions from AGB stars in these populations. This could also indicate the presence of Ba-rich stars resulting from binary processes. Higher AGB contributions imply these populations underwent prolonged or multiple-phase star formation, allowing intermediate-mass stars to evolve into AGB stars, enriching the ISM with s-process elements. The `spray' has a varied range of abundance ratios for the other elements considered, showing no clear pattern like the `core' or `branch', which makes it a more heterogeneous group in terms of chemical composition.

\item Branch: The `branch' contains more metal-poor stars, with [Fe/H] $\approx -1.4$ dex. It exhibits significant variance in [Eu/Fe], pointing to stochasticity in r-process element production in the metal-poor regime. The `branch' exhibits lower [Al/Fe] than the `core' or `spray' regions, indicating significant differences in SFHs, with low [Al/Fe] being a property of accreted populations. Many of the iron-peak elements have negative weights in PC1, which can be seen through the sub-solar values of elements such as [Cu/Fe] towards the `branch'. A similar trend is observed in [Cu/Fe], where the lower [Cu/Fe] agrees with findings for accreted halo stars. The `branch' also has high [Ba/Fe] values, and a slightly sub-solar [Mg/Fe] present in the lower region, potentially indicative of low-$\alpha$ halo stars. In PC1, [Ba/Fe], [Zn/Fe], [Y/Fe], and [La/Fe] have positive weights, indicating these elements increase towards the accreted population in the `branch'.

\end{itemize}

The PC1-PC2 hyperplane is just a two-dimensional representation of the reduced $9$-dimensional $\mathcal{C}$-space, which indicates why there are only three components selected by eye (see Figure \ref{fig:PCA_labelled}). This will be explored in more detail in the following Section \ref{sec:cluster_analysis}, where we cluster the latent space using XD to estimate the underlying number of groups.

\subsection{Clustering the PCA-Transformed Space}\label{sec:cluster_analysis}

\begin{figure}
 \centering
    \includegraphics[width=\linewidth]{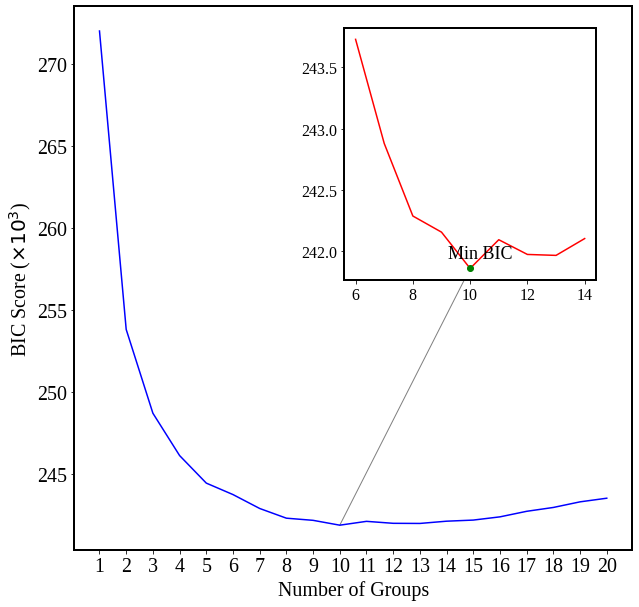}
 \caption{BIC scores for clustering the $9$D-transformed $\mathcal{C}$-space of metal-poor MW stars. The minimum BIC, the `knee', suggests $10$ groups. The inset shows the minimum BIC score at $10$ groups. Although $10$ to $14$ groups yield similar scores, we proceed with $10$ groups for simplicity and robustness.}
 \label{fig:BIC_PCA_9D}
\end{figure}

Our goal is to cluster the metal-poor MW stars into groups with similar chemistry to better understand the chemical diversity and formation history of components such as the thick disc, thin disc, and stellar halo. We apply XD to our $9$D-transformed $\mathcal{C}$-space for a varying number of groups and calculate the BIC for each case. The BIC scores shown in Figure \ref{fig:BIC_PCA_9D} suggest the presence of $\approx 10$ groups, which we label as G1-G10. It is important to note that the optimal number of groups is not very clear from the BIC analysis, as a range of about $10$ to $14$ groups returns similar BIC scores. For simplicity and to facilitate our analysis, we proceed with the minimum number of groups suggested by the BIC knee, which is $10$, noting also that a slight under-estimation of the total number of groups is preferable to an over-estimation, as discussed in Section \ref{sec:group_number}. Additionally, we only assign stars to a group if they have a $\geq 70\%$ probability of belonging to that Gaussian component.

\subsection{Mapping the Groups to Known Populations}\label{sec:origins}

\begin{figure*}
  \centering
  \baselineskip=0pt
    \subfigure{
      \includegraphics[width=0.99\linewidth]{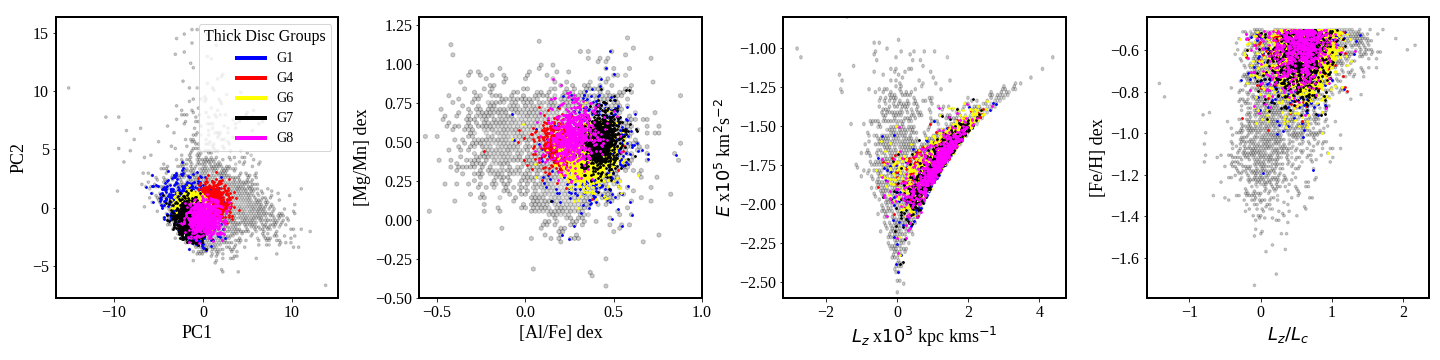}
    }\vspace{-3mm}  
    \subfigure{
      \includegraphics[width=0.99\linewidth]{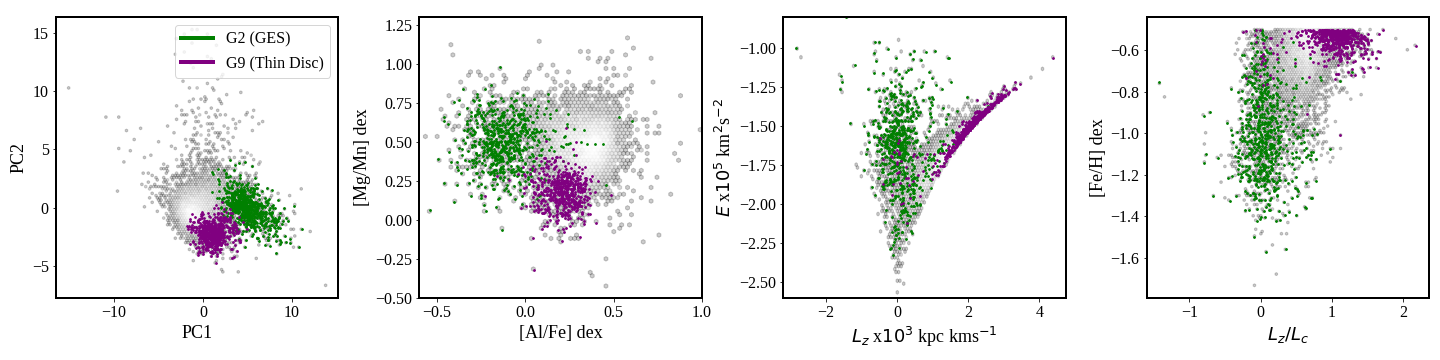}
    }\vspace{-3mm}  
    \subfigure{
      \includegraphics[width=0.99\linewidth]{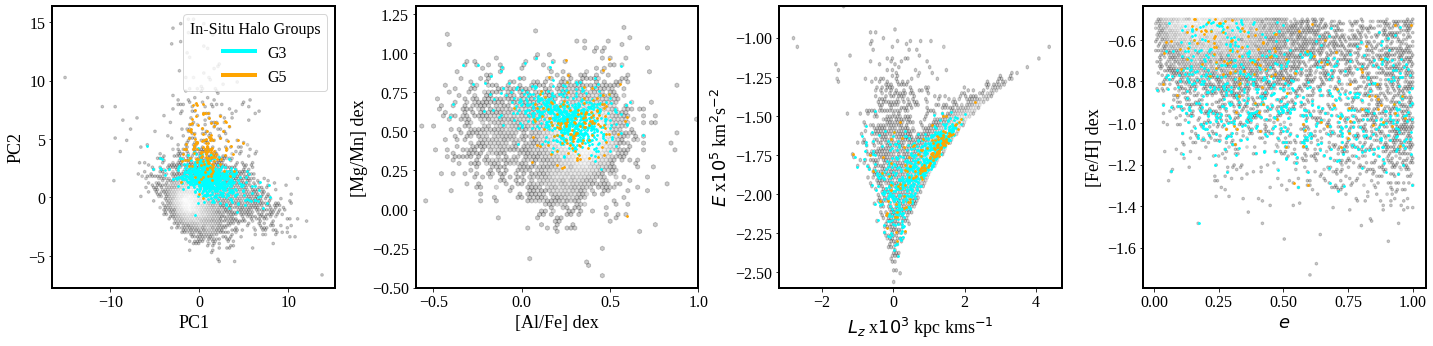}
    }
  \caption{Plots of all groups in varying chemodynamical and PC planes. Top: the thick disc groups, G1, G4, G6, G7 and G8, in order from left to right: PC2 - PC1, [Mg/Mn] - [Al/Fe], $L_{z}$ - $E$ and $L_{z}/L_{c}$ - [Fe/H]. Middle: GES (G2) and the thin disc (G9), in order from left to right: PC2 - PC1, [Mg/Mn] - [Al/Fe], $E$ - $L_{z}$ and [Fe/H] - $L_{z}/L_{c}$. Bottom: the in-situ halo groups (G3 and G5) in order from left to right: PC2 - PC1, [Mg/Mn] - [Al/Fe], $E$ - $L_{z}$ and [Fe/H] - $e$.}
  \label{fig:Group_chemodynamical}
\end{figure*}

In Figure \ref{fig:Group_chemodynamical}, we show the location of the groups in commonly used chemodynamical hyperplanes, [Mg/Mn] - [Al/Fe], $L_{z}$ - $E$ and $L_{z}/L_{c}$ - [Fe/H], while also showing where the groups lie in the reduced PC1 - PC2 space. While [Mg/Fe] has a relatively low weight of $-0.2$ in PC1 (shown in Figure \ref{fig:PC1_PC2}), indicating it does not vary significantly, we still utilise it to explore its associations with the thick disc, thin disc and halo populations. Combining this with information from the other [X/Fe] distributions for each group, found in the Appendix \ref{sec:appendix}, we can determine the chemodynamical properties of each of them. We also refer the reader to Table \ref{tab:group_characteristics} for a complete collection of means and errors of the chemodynamical distributions.

\subsubsection{G1, G4, G6, G7 and G8 - Thick Disc Groups}

G1, G4, G6, G7, and G8 contain $763$, $648$, $2235$, $832$, and $610$ stars, respectively. In the PC1-PC2 plane these groups form the `core'. They occupy an $\alpha$-rich region with [Mg/Fe]$_{\mu}$ values of $\approx 0.3$ dex. Their [Fe/H] distributions show the most metal-enhanced values among the groups of around $-0.6$ dex. All have prograde orbits ($L_{z} \approx 1200$ kpc kms$^{-1}$) and $e$ distributions peaking at $\approx e = 0.30$. 

Regarding their chemistry, the [Ba/Fe] values differ: G4 and G8 have super-solar values ([Ba/Fe]$_{\mu} = 0.28$ dex and [Ba/Fe]$_{\mu} = 0.21$ dex respectively), while G6 and G7 have slightly sub-solar values ([Ba/Fe]$_{\mu} = -0.05$ dex and [Ba/Fe]$_{\mu} = -0.10$ dex respectively). The [Y/Fe] trend is similar. [Eu/Fe] abundances are consistently super-solar across these groups ($\approx 0.3$ dex). G4 shows a high Zn enhancement ([Zn/Fe]$_{\mu} = 0.42$ dex), while others are closer to solar ($\approx 0.1$ dex). G7 and G8 show bi-modalities in [Zn/Fe] with solar and sub-solar components (e.g., G7$_{\mathrm{subsolar}} = -0.3$ dex and G7$_{\mathrm{solar}} = 0.1$ dex). The [Al/Fe] distribution is narrow across all groups, with abundances ranging from [Al/Fe]$_{\mu} = 0.22 - 0.45$ dex. For [Cu/Fe], all groups have solar values except G4, which has [Cu/Fe]$_{\mu} = -0.08$ dex. In [Ni/Fe], all groups have super-solar Ni-abundances except G4 ([Ni/Fe]$_{\mu} = 0.00$ dex). Finally, the [Sc/Fe] distributions are similar across G4, G6, and G7, with tails extending between [Sc/Fe] $= 0.0 - 0.2$ dex. G1 has an extended distribution with [Sc/Fe] reaching $\approx 0.4$ dex. G8 is the only group with sub-solar values, [Sc/Fe]$_{\mu} = 0.04$ dex, and a tail extending down to $\approx -0.2$ dex. These groups (G1, G4, G6, G7, and G8) lie in the thick disc region of the [Mg/Mn] - [Al/Fe] plane.

The properties of these groups, including their $\alpha$-rich nature, relatively high metallicity, prograde orbits, and specific chemical abundance patterns, align with the expected SFH of the thick disc. This history involves rapid early star formation from gas enriched by CCSNe, followed by prolonged or multiple-phase star formation episodes \citep{bland-hawthornGalaxyContextStructural2016, yuBurstyOriginMilky2021}. G4 and G8 have some exceptions to expected abundance patterns, in the case of [Ba/Fe] and [Zn/Fe], this is further discussed in Section \ref{sec:Breakdown of the thick disc}.

\subsubsection{G2 - GES}

G2 contains $671$ stars and dominates the `branch' in the PC2 - PC1 plane. G2 has low-metallicity values ([Fe/H]$_{\mu} = -1.00$ dex) and $L_{z,\mu} = 145.13 \pm 521.97$ kpc kms$^{-1}$, with an eccentricity $e_{\mu} = 0.80 \pm 0.21$, extending up to a peak at $e \approx 1$, making it the most dynamically hot of all groups. 

It is $\alpha$-rich, slightly less so than the groups in the ‘core’, with [Mg/Fe]$_{\mu} = 0.13$ dex, and is the most Mn-depleted group present, with [Mn/Fe]$_{\mu} = -0.37$ dex. G2 has a high Ba enhancement, [Ba/Fe]$_{\mu} = 0.41$ dex, though the ls-process element Y enhancement is [Y/Fe]$_{\mu} = 0.11$ dex, indicating a lower abundance compared to the hs-process. G2 shows super-solar [Eu/Fe] values at $\mu = 0.44$ dex and a uni-modal, solar [Zn/Fe] distribution with [Zn/Fe]$_{\mu} = 0.08$ dex. The [Al/Fe] distribution is sub-solar, the lowest across all groups, with [Al/Fe]$_{\mu} = -0.12$ dex. Notably, G2 has a dominant peak at very sub-solar [Cu/Fe] values, [Cu/Fe]$_{\mu} = -0.46$ dex, making it the least Cu-enriched in the sample. The [Ti/Fe] distribution shows that while G2 is enhanced in Mg (which is also an $\alpha$-element), it extends to [Ti/Fe] abundances as low as those of G9. G2 has [Ni/Fe]$_{\mu} = -0.11$ dex, and [Sc/Fe]$_{\mu} = -0.01$ dex. 

In the [Mg/Mn] - [Al/Fe] plane, G2 overlaps with the `blob’ identified in \citet{dasAgesKinematicsChemically2020} and also mirrored by other studies \citep{myeongMilkyWayEccentric2022}, indicating the presence of GES stars. The low metallicity of these stars position G2 within the more metal-rich extent of GES ([Fe/H]$_{\mu}= -1.05 \pm 0.20$ dex \citep{ortigoza-urdanetaGalacticArchaeoLogIcaLExcavatiOns2023}, although GES is thought to extend down to [Fe/H] $\approx -1.5$ dex \citep{naiduEvidenceDisruptedHalo2022}, a metal-poor region not fully captured in our sample, which extends down to [Fe/H] $\approx -0.6$ dex. G3 and G10 also show GES-like stars, with G10 reaching lower metallicities. The high [Ba/Fe] values in G2 present a discrepancy with studies that indicate GES has negligible s-process contribution and sub-solar [Ba/Fe] values ($-0.02$ dex) \citep{carrilloDetailedChemicalAbundance2022}. However, our [Ba/Fe]${\mu} = 0.41 \pm 0.19$ dex closely matches \citet{myeongMilkyWayEccentric2022}’s estimate for metal-rich GES ([Ba/Fe] = $0.38$ dex, [Fe/H] $= -0.96$ dex). G2 also shows r-process enhancement with [Eu/Fe]$_{\mu} = 0.44$ dex, aligning with the metal-rich GES cited in the literature ([Eu/Fe] $= 0.47$ dex \citep{myeongMilkyWayEccentric2022}). The sub-solar [Al/Fe]$_{\mu} = -0.12$ dex is consistent with an accreted population, matching \citet{myeongMilkyWayEccentric2022}’s metal-rich GES ([Al/Fe] $= -0.17 \pm 0.09$ dex). G2’s extreme sub-solar [Cu/Fe] values define it distinctly. 

\subsubsection{G9 - Thin Disc}

G9 contains $581$ stars and is part of a distinct lower cluster in the segmented `core' shown in Figure \ref{fig:PCA_labelled}. G9 has the highest $L_{z}$ among the groups, with $L_{z} = 2020.10 \pm 619.82$ kpc kms$^{-1}$. There is a slight bi-modality in $L_{z}$, with a small group of stars having $L_{z} \sim 0$, and it has the most circular orbits of all the groups, with a distinct peak at $e \approx 0.1$. G9 has a tail towards $L_{z} \leq 0$, with $9$ stars exhibiting retrograde kinematics, representing $\approx 1.72\%$ contamination from G2. G9 is the most $\alpha$-poor group in the sample, with [Mg/Fe]$_{\mu} = 0.10$ dex, supported by low [Ti/Fe] values ([Ti/Fe]$_{\mu} = 0.09$ dex). It has the most overall Mn-rich distribution of the groups, with [Mn/Fe]$_{\mu} = -0.08$ dex), with only G1 and G10 showing extended tails to higher Mn-enhancement. The [Fe/H] distribution is the narrowest and also among the most metal-rich ([Fe/H]$_{\mu} = -0.56$ dex). The s-process enhancement shows slight bi-modality, more evident in the ls-process element Y, with [Y/Fe] featuring peaks at solar ($\approx 0.1$ dex) and sub-solar ($\approx -0.2$ dex) values. G9 has the lowest r-process enhancement with [Eu/Fe]$_{\mu} = 0.15$ dex). The [Zn/Fe] values are solar, [Zn/Fe]$_{\mu} = -0.01$ dex, and feature a tail extending to low [Zn/Fe] values ($< -0.5$ dex). [Al/Fe] aligns with the ‘core’ groups, and the [Cu/Fe] distribution closely resembles that of G1, G6, and G8. G9 has the lowest [Ti/Fe] of all groups, [Ti/Fe]$_{\mu} = 0.09$ dex, with a tail extending to sub-solar values, $\approx -0.2$ dex, similar to G2. The [Ni/Fe] distribution is similar to G1, G4, G6, and G7 ([Ni/Fe]$_{\mu} = 0.04$ dex), and [Sc/Fe] values closely match G8, at [Sc/Fe]$_{\mu} = 0.06$ dex. In the [Mg/Mn] - [Al/Fe] plane, G9 lies in a region characteristic of the thin disc. This is corroborated by its predominantly prograde and circular orbits. The $\alpha$-poor nature and narrow metallicity distribution at [Fe/H] $> -0.8$ dex are consistent with the thin disc population studied by \citet{hawkinsUsingChemicalTagging2015}. The retrograde contamination ($1.72\%$) likely originates from GES, as they possess a similar $\alpha$ enhancement. 

\subsubsection{G3 and G5 - In-situ Halo}

G3 and G5 contain $651$ and $141$ stars, respectively. They both occupy the upper region of the `core' (at PC1 $\approx 0$ and PC2 $\approx 5$), with G3 extending into the `branch' and G5 extending into the `spray'. It is evident that these two groups overlap across multiple hyperplanes, capturing stars on retrograde and more circular orbits. G3 features a bi-modality at $L_{z} \approx 1000$ and $0$ kpc kms$^{-1}$. The prograde peak aligns with G1, G4, G6, G7, and G8 (i.e. the thick-disc populations), while the $L_{z} \approx 0$ kpc kms$^{-1}$ peak matches G2 (GES). G5 also shows a bi-modal distribution in $L_{z}$, with stars at $L_{z} = 0$ and retrograde values. The $e$ distribution in both groups is widely spread between $e = 0$ and $1$, with a bi-modality of thick disc-like and higher eccentricity orbits.

The [Mg/Fe] distribution of G3 is centered at [Mg/Fe]$_{\mu} = 0.32$ dex, similar to the $\alpha$-enhanced G1, G4, G6, G7, and G8 (thick-disc populations). G5 has the most $\alpha$-rich stars of all groups, with [Mg/Fe]$_{\mu} = 0.36$ dex, and a distribution tail extending to $\approx 0.75$ dex. Both groups have wide metallicity distributions, with G5 being more metal-rich ([Fe/H]$_{\mu} = -0.75$ dex), and G3 having a lower metallicity ([Fe/H]$_{\mu} = -0.87$ dex). The Ba-enrichment of G3 is similar to G2 ([Ba/Fe]$_{\mu} = 0.44 \pm 0.24$ dex), while G5 has the highest [Ba/Fe] enrichment among all groups ([Ba/Fe]$_{\mu} = 0.97 \pm 0.37$ dex). G3 has a narrow [Y/Fe] distribution ([Y/Fe]$_{\mu} = 0.33 \pm 0.17$ dex), unique among the groups, while G5 has a higher enhancement, [Y/Fe]$_{\mu} = 0.68 \pm 0.26$ dex, with a complex distribution featuring about four peaks. The high s-process enhancements present in G3 are found in the proto-galaxy population Aurora \citep{myeongMilkyWayEccentric2022}, and are attributed to the prolonged star formation of in-situ populations. Similarly, Heracles \citep{hortaEvidenceAPOGEEPresence2020, naiduEvidenceDisruptedHalo2022}, an ancient merger event prior to the GES merger has similar s-process properties to G3. The Eu-enhancement of G3 closely resembles G1 and G6 (thick-disc groups), featuring a tail of Eu-enhanced stars trailing up to [Eu/Fe] $> 0.7$ dex. G5 has a similar distribution but with a slightly higher mean enhancement ([Eu/Fe]$_{\mu} = 0.39$ dex), and a tail extending up to [Eu/Fe] $\approx 0.8$ dex. 

The [Al/Fe] distribution are $0.25$ and $0.31$ dex for G3 and G5 respectively, and are too high to be considered from Heracles, which is limited to [Al/Fe] $< 0$ \citep{hortaEvidenceAPOGEEPresence2020}. This range however matches well with the in-situ halo population Aurora, with [Al/Fe] $= 0.10 \pm 0.16$ dex \citep{myeongMilkyWayEccentric2022}. For [Zn/Fe], G3 has [Zn/Fe]$_{\mu} = 0.19$ dex and G5 has [Zn/Fe]$_{\mu} = 0.25$ dex, similar to G1, G6, and G8. Both G3 and G5 extend to the lowest [Cu/Fe] values ($\approx -0.5$ dex), with G3 having the lowest mean ([Cu/Fe]$_{\mu} = -0.14$ dex). The [Ti/Fe] distributions for G3 and G5 are similar to G1, G6, and G7 (thick-disc groups), with [Ti/Fe]$_{\mu} = 0.26$ dex and [Ti/Fe]$_{\mu} = 0.28$ dex respectively, which is also seen in the case of [Ni/Fe]. The [Sc/Fe] distributions of both G3 and G5 closely match G9 and G8, with [Sc/Fe]$_{\mu} = 0.08$ dex and [Sc/Fe]$_{\mu} = 0.11$ dex respectively. Retrograde contamination in these groups is $99$ stars for G3 ($15.2\%$ of the group) and $15$ stars for G5 ($10.6\%$ of the group). Located in the [Mg/Mn] - [Al/Fe] plane where in-situ halo stars are found, G3 and G5 are distinguished by their s-process enhancement in [Ba/Fe] and [Y/Fe].

\subsubsection{G10 - Outlier Group}

G10 contains $365$ stars and is an outlier group, overlapping into the `branch’, `spray’, and `core’. G10 appears as a background component in all hyperplanes. In $L_{z}$, G10 has two peaks with a greater spread towards prograde values ($L_{z} > 2000$ kpc kms$^{-1}$), and a wide $e$ distribution. G10 shows a broad range of [Mg/Fe] values from $-0.75$ to $0.75$ dex, grouping together some of the most $\alpha$-rich and $\alpha$-poor stars in the data set. It also has a spread of extreme [Ba/Fe] values, likely captured in the `spray’, with hs-process enhancement ([Ba/Fe] $> 2.0$ dex). [Y/Fe] shows a similar spread. The [Eu/Fe] distribution has wide dispersion and tails to very high Eu-enhancements ([Eu/Fe] $\approx 1.0$ dex). In contrast, [Ni/Fe] and [Sc/Fe] distributions are narrower than other chemical abundances, with [Ni/Fe]$_{\mu} = 0.00 \pm 0.16$ dex and [Sc/Fe]$_{\mu} = 0.12 \pm 0.17$ dex, showing similarities to G5 in both abundances. Kinematically, G10 includes approximately two distinct groups that align with the thick disc kinematics of G1, G4, G6, G7, and G8, as well as the accreted, $L_{z} \approx 0$, G2 group. The wide distribution of chemical and kinematic features contrasts with its tight distributions in both [Ni/Fe] and [Sc/Fe].

\section{Discussion}\label{sec:discussion}

In this section, we assess the robustness of our method and compare our findings with previous studies. We discuss the chemical and kinematic properties of different groups, emphasising the distinct populations within the thick disc and their implications for the SFH of the metal-poor MW. Additionally, we examine the in-situ halo groups and GES and explore the chemical diversity present in our sample.

\begin{figure*}
 \centering
    \includegraphics[width=0.8\linewidth]{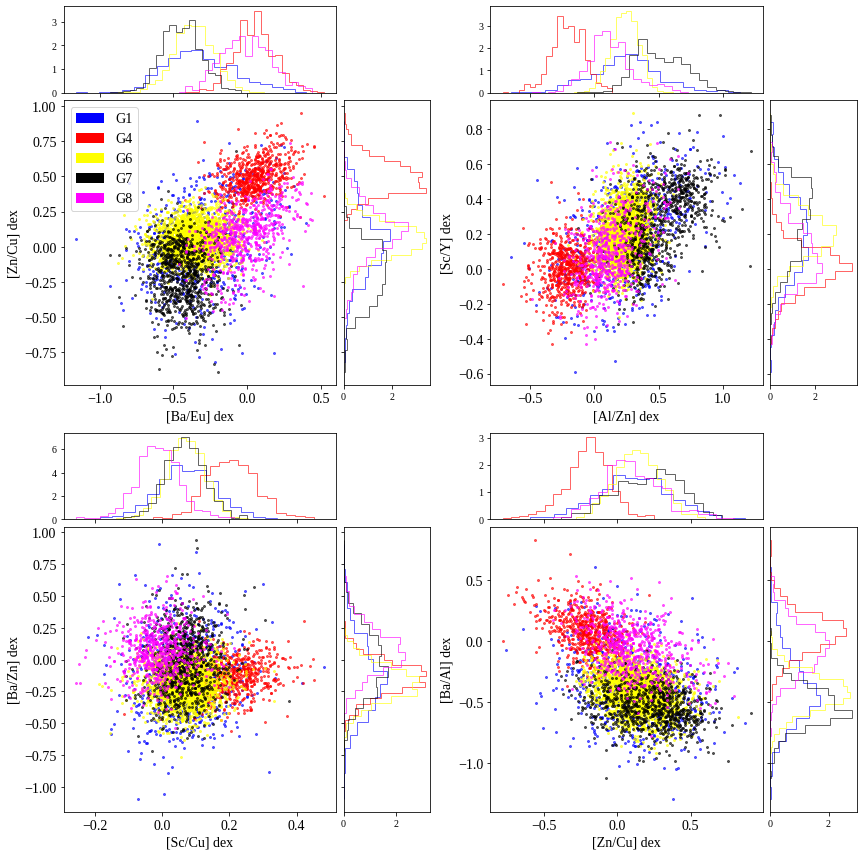}
 \caption{Four hyperplanes: (left to right, top to bottom) [Zn/Cu] - [Ba/Eu], [Sc/Y] - [Al/Zn], [Ba/Zn] - [Sc/Cu] and [Ba/Al] - [Cu/Y]. The five groups are selected to belong to the thick disc as defined by their high [$\alpha$/Fe] and thick disc-like kinematics. In these chemical hyperplanes, the thick-disc populations show distinction in chemistry. G4 is easier to classify separately in each plane, whereas G7 is most clearly distinguishable in [Al/Zn] and [Zn/Cu].}
 \label{fig:thick_decompose}
\end{figure*}

\subsection{Multiple Populations in the Thick Disc} \label{sec:Breakdown of the thick disc}

The magnitude of the weights for Ba (PC1 and PC2), Y (PC2), Zn (PC2), Al (PC1), Cu (PC1), La (PC2), Sc (PC2) and Eu (PC2) are among the highest, and the distinguishing power of these chemical abundances become a useful tool to decompose the thick disc. Figure \ref{fig:thick_decompose} shows the chemical composition of groups we associate with the thick disc. We select the following hyperplanes: [Zn/Cu] - [Ba/Eu], [Sc/Y] - [Al/Zn], [Ba/Zn] - [Sc/Cu] and [Ba/Al] - [Cu/Y]. [Zn/Cu] is selected as they both have different timescales: Cu from SNe1a and AGB (longer delay) and Zn from CCSNe and HNe (shorter delay). Hence, regions with early star formation might have higher [Zn/Cu], while prolonged star formation shows lower [Zn/Cu]. [Ba/Eu] traces s- vs r-process nucleosynthesis. Higher [Ba/Eu] indicates a stronger AGB star influence and prolonged star formation; lower [Ba/Eu] suggests significant r-process events linked to early, rapid star formation. Thus in Figure \ref{fig:thick_decompose}, G4 in the top right (high [Zn/Cu], high [Ba/Eu]) could represent an older population with rapid star formation and early enrichment. We note that G4 lies in the region of the chemical hyperplane, and suspect it could have experienced a high number of HNe. [Sc/Y] indicates the play-off between CCSNe and the ls-process in AGB stars- Y shows a very high weighting in PC2 and this variance is seen in [Sc/Y]. [Al/Zn] is another probe for HNe, as well as for lower mass (low Al) populations. This means G4 and G8 both have low [Al/Zn], indicating lower mass systems with evidence of early intense star formation. The bottom two chemical hyperplanes also demonstrate how the thick disc populations lie, with G4 and G8 being the most distinctly separate. 

Overall, G1 and G6 are chemically similar, representing the classical thick disc. G7 differs due to its Mn and Zn enhancement, while G4 is the most chemically distinct among the thick disc groups captured here. G8 alongside G4 has a distinctive hs-element enhancement which is captured in [Ba/Fe]. This enhancement in Ba, but not Y, could indicate that there was an environment with higher neutron density \citep{bisterzoProcessLowmetallicityStars2010}. Additionally, G4 features a very high [Zn/Fe] enhancement which could be the contribution from HNe \citep{kobayashiOriginElementsCarbon2020}. This suggests while some studies have argued that the thick disc is mostly chemically homogeneous \citep{haywoodRevisitingLongstandingPuzzles2019, chandraThreePhaseEvolutionMilky2023}, we find evidence that the thick disc's formation could have resulted from three separate events or nuanced nucleosynthesis channel which results in the classical thick disc (G1 + G6 + G7), G4 (high Zn, high Ba) and G8 (low Sc, high Ba) \citep{navarroThickThinKinematic2011, belokurovDawnTillDisc2022}.

\subsection{In-situ halo and GES}

G3 and G5 are likely the in-situ halo. We look to Heracles, postulated to be an ancient accreted population. The stars are proposed to have formed prior to the spin-up of the MW disc. It was discovered chemodynamically by \citet{hortaEvidenceAPOGEEPresence2020} and is thought to have an estimated progenitor stellar mass of $M_{*} \approx 5 \times 10^{8} M_{\odot}$. The [Al/Fe] distribution presented in this paper is on average higher than the expectation for Heracles quoted in \citet{hortaChemicalCharacterizationHalo2023}. This is because the authors selected Heracles according to the following chemical criteria: [Al/Fe] $< -0.07 \&$ [Mg/Mn] $\geq 0.25$, and [Al/Fe] $\geq -0.07 \&$ [Mg/Mn] $\geq 4.25 \times$ [Al/Fe] $+ 0.5475$. Other connections have been made relating to this population, e.g. Kraken \citep{kruijssenKrakenRevealsItself2020} and Koala \citep{forbesReverseEngineeringMilky2020}.

Heracles has been found to have very similar properties to the proto-galaxy population, Aurora, as identified in \citet{myeongMilkyWayEccentric2022}. Aurora is associated with having a higher [$\alpha$/Fe] ratio compared with Heracles, indicating that Aurora had a higher SFR than Heracles \citep{hortaChemicalCharacterizationHalo2023}. Additionally, as G3 occupies a lower orbital energy region, $E_{\mu} = -1.82 \times 10^{5}$ km$^{2}$s$^{-2}$, this appears to kinematically match with Aurora which has orbital energies of $-1.76 \pm 0.15 \times 10^5$ km$^{2}$s$^{-2}$. Additionally, as noted, G3 is incredibly enhanced in the [Ba/Fe] and [Y/Fe] ratio, indicating the contribution of metallicity-dependent AGB stars. Our values of s-process enhancement, [Ba/Fe] $= 0.44 \pm 0.24$ dex, align with those presented for Aurora at [Ba/Fe] $= 0.48 \pm 0.25$ dex \citep{myeongMilkyWayEccentric2022}.

G2's chemodynamical properties point to it being representative of the accreted population GES. This structure has been extensively studied in terms of its chemodynamical properties. In this work, we find that G2 has lower $\alpha$-enrichment than the thick disc, in agreement with previous works (e.g. \citep{haywoodDisguiseOutReach2018, helmiMergerThatLed2018, mackerethOriginAccretedStellar2019, dasAgesKinematicsChemically2020, buderGALAHSurveyChemical2022, hortaChemicalCharacterizationHalo2023}). Additionally, as GES is an accreted population we find that it has a sub-solar [Al/Fe] ratio which resembles the abundance pattern of stars from satellites of the MW \citep{hawkinsUsingChemicalTagging2015, hortaEvidenceAPOGEEPresence2020}. Looking to the neutron-capture elements, we see that G2 has a high [Eu/Fe] ratio of $0.44 \pm 0.15$ dex, this aligns with many other sources (see e.g. \citep{matsunoProcessEnhancementsGaiaEnceladus2021, aguadoElevatedRprocessEnrichment2021, naiduEvidenceDisruptedHalo2022, myeongMilkyWayEccentric2022, carrilloDetailedChemicalAbundance2022}). On the other hand, looking to the s-process abundances, such as Ba and Y, we see that our sample of GES candidates have an equivalent enhancement to Eu ([Ba/Fe] $=0.41 \pm 0.19$ dex). This is in contradiction with some studies who indicate that GES has solar levels of Ba-enhancement (e.g. \citep{carrilloDetailedChemicalAbundance2022}), while others measure a strong enhancement of up to [Ba/Fe] $= 0.38 \pm 0.18$ dex \citep{myeongMilkyWayEccentric2022}. GES is thought to have an extended SFH of $\approx 3.6$ Gyr \citep{bonacaOrbitalClusteringIdentifies2021}, meaning theoretically there would be an expected source of enrichment via AGB stars.

\subsection{Method Robustness and Comparison}

We find in this study of observational data that $9$ PCs are required to make up $85\%$ of the cumulative variance, reflecting the findings of \citet{tingPrincipalComponentAnalysis2012} and \citet{price-jonesBlindChemicalTagging2019a}, who argued for the necessity of $\approx 6 - 10$ abundance dimensions. The results contrast however with those obtained from applying PCA to a set of chemical abundances predicted by a one-zone chemical evolution model, \citet{andrewsInflowOutflowYields2017}. Here, the authors found that $\alpha$-elements and elements such as Fe, Mn, and Al dominate the encoded information, with two PCs accounting for $99\%$ of the variance in their model. This suggests that chemical evolution models may be underestimating the true diversity of contributions to the high-dimensional $\mathcal{C}$-space. While we find that Ba, Zn, and Cu are all important elements in chemically distinguishing stars, we do not see a significant variance in [Mg/Fe]. \citet{tingPrincipalComponentAnalysis2012} found that low-mass AGB stars are significant contributors to the abundance of hs and ls-elements at higher metallicity, which could reflect the Ba-enhancement we see in G4 and G8 among the thick-disc groups. The s-process elements are also shown to be effective for chemical tagging in \citet{blanco-cuaresmaTestingChemicalTagging2015}. \citet{weinbergChemicalCartographyAPOGEE2022} emphasises that to effectively chemically tag groups, it is recommended to minimise group overlap and reduce residual correlations between abundances to below the level of observational uncertainties.

\begin{figure}
  \centering
    \includegraphics[width=0.6\linewidth]{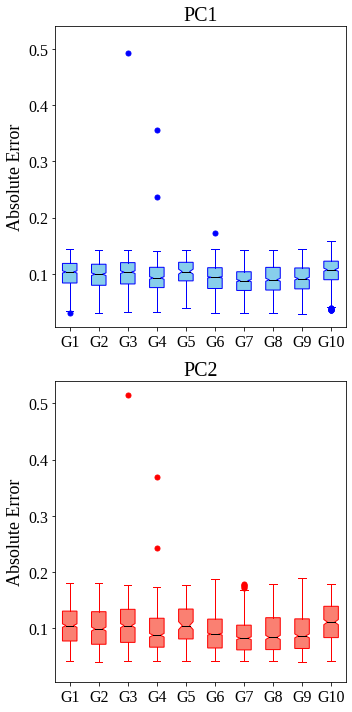}
  \caption{These box plots show the spread and distribution of absolute errors in PC1 (upper) and PC2 (lower) for all the groups. The whiskers extend to 1.5 $\times$ the interquartile range. Data points beyond the whiskers are considered outliers and are marked as individual dots. It's unlikely that G10's wide distribution of chemodynamical properties are due to large errors, as shown.}
  \label{fig:G10_Errors}
\end{figure}

\citet{hoggCHEMICALTAGGINGCAN2016} assessed the potential of chemical tagging in light of challenges such as noise and incompleteness in chemical abundance measurements and determined that clustering in an abundance space can be effective. We also find that by using the PCA latent space as the clustering target we are able to cleanly recover different histories in the metal-poor MW for the majority of groups. In \citet{naiduEvidenceH3Survey2020}, they conclude that clustering algorithms (such as DBSCAN, HBDSCAN and k-means) either fracture the space into too many clusters, or assign all stars to a dominant structure (e.g. GES or the thick disc). Furthermore, \citet{blanco-cuaresmaTestingChemicalTagging2015} indicate that chemical homogeneity across differing groups becomes a challenge, with there being a non-negligible overlap between structures dependent on the $\mathcal{C}$-space. In light of this, we will discuss the outlier group (G10). The errors of the PC1 and PC2 variables for each group are shown in Figure \ref{fig:G10_Errors}, indicating that high errors are not the reason for this group capturing an extreme range of chemodynamical properties.

\begin{figure*}
  \centering
  \includegraphics[width=\linewidth]{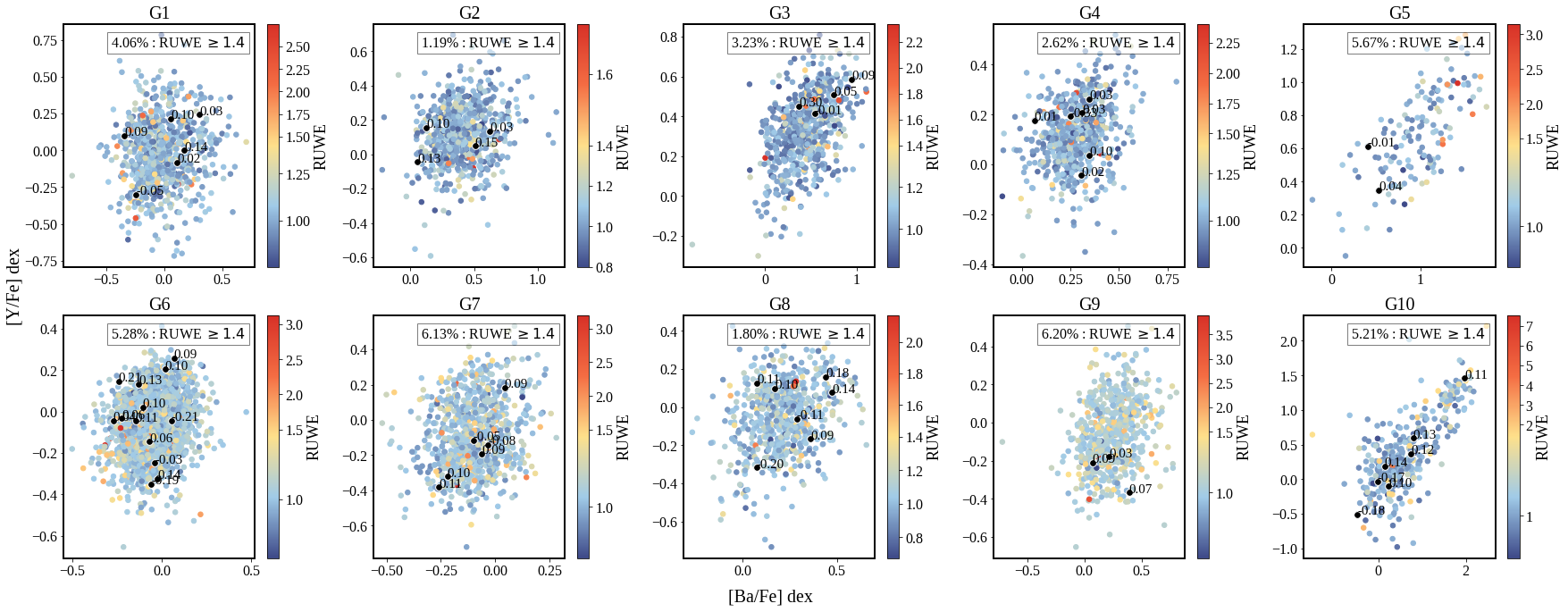}
  \caption{G1 to G10 plotted in the [Y/Fe] - [Ba/Fe] chemical space, and colour-mapped by the RUWE values. At RUWE $\geq 1.4$ \citep{buderGALAHSurveyThird2021}, the single-star astrometric fit becomes poor and this could indicate the presence of an unresolved binary. The number of stars with RUWE $\geq 1.4$ is very low across all groups. Black points represent stars with [C/Fe] values obtained from cross-matching APOGEE and GALAH data.}
  \label{fig:RUWE}
\end{figure*}

Also, we look at the Renormalised Unit Weight Error (RUWE) values of the s-process enhanced in-situ halo groups (G3 and G5) to evaluate whether they are unresolved AGB binaries which may possess enhanced s-process values due to mass transfer and other effects. After examining the RUWE values across all groups, we quantified the number with RUWE $\geq 1.4$, based on the recommended cut-off value from \citet{buderGALAHSurveyThird2021}. The number of stars with RUWE $\geq 1.4$ is very low across all groups, indicating that unresolved binaries are not a significant concern. Additionally, we cross-matched APOGEE \citep{prietoAPOGEEApachePoint2008} and GALAH surveys to obtain [C/Fe] values for $329$ stars, as high C along with Ba is another indicator of binary AGB stars. The data suggest that Ba-rich stars are not a concern in our analysis. This is shown in Figure \ref{fig:RUWE}.

\section{Conclusions}

In this study, we explored the chemical diversity of the MW's metal-poor populations by combining PCA and XD on $17$ abundances measured for $9,923$ stars in the GALAH DR3 dataset. This methodology allowed for an unbiased and comprehensive examination of the Galaxy's chemical landscape, leading to the identification of $10$ distinct stellar groups. Our approach provides a clear advantage over traditional techniques by avoiding arbitrary cuts based on chemistry or kinematics, offering a less biased representation of the Galaxy's chemical complexity. Our key findings include:

\begin{itemize}

\item Identification of Known Populations: By using chemistry alone, we successfully identified known stellar populations through a combination of latent space representation and a detailed examination of chemical abundances beyond a traditional 2D analysis. This includes GES, the thick disc, the thin disc and in-situ halo populations.

\item Limited Discriminatory Power of Mg: Mg has less impact in separating accreted and disc stars, with the greatest $\alpha$-separation being between the thick and thin discs. Other elements like Ba, Al, Cu, and Sc are important for distinguishing disc from accreted stars, and Ba, Y, Eu, and Zn for distinguishing disc and accreted stars from in situ halo.

\item Multiple Chemically Distinct Populations in the Thick Disc: The thick disc is revealed to be a composite of multiple populations with distinct chemical histories, underscoring its complexity and the consideration of it being a diverse assembly rather than a homogeneous entity.

\end{itemize}

These findings highlight the intricate chemical structure of the thick disc and the broad range of influential elements, reflecting the diverse enrichment environments specific to the metal-poor MW.

\section*{Acknowledgements}

PD is supported by a UKRI Future Leaders Fellowship (grant reference MR/S032223/1). NB
would like to thank the GLEAM and ERIS collaborations for supporting this research. The research for this publication was coded in \verb|python| and included its packages: \verb|numpy|, \verb|astropy|, \verb|pandas|, \verb|matplotlib| and \verb|sklearn|. This work made use of the \verb|Extreme Deconvolution| (XD) package \citep{bovyExtremeDeconvolutionInferring2011}, available at: https://github.com/jobovy/extreme-deconvolution. This work made use of the Third Data Release of the GALAH Survey \citep{buderGALAHSurveyThird2021}. The GALAH Survey is based on data acquired through the Australian Astronomical Observatory, under programs: A/2013B/13 (The GALAH pilot survey); A/2014A/25, A/2015A/19, A2017A/18 (The GALAH survey phase 1); A2018A/18 (Open clusters with HERMES); A2019A/1 (Hierarchical star formation in Ori OB1); A2019A/15 (The GALAH survey phase 2); A/2015B/19, A/2016A/22, A/2016B/10, A/2017B/16, A/2018B/15 (The HERMES-TESS program); and A/2015A/3, A/2015B/1, A/2015B/19, A/2016A/22, A/2016B/12, A/2017A/14 (The HERMES K2-follow-up program). We acknowledge the traditional owners of the land on which the AAT stands, the Gamilaraay people, and pay our respects to elders past and present. This paper includes data that has been provided by AAO Data Central (datacentral.org.au). This work has made use of data from the European Space Agency (ESA) mission Gaia (https://www.cosmos.esa.int/gaia), processed by the Gaia Data Processing and Analysis Consortium (DPAC, https://www.cosmos.esa.int/web/gaia/dpac/consortium). Funding for the DPAC has been provided by national institutions, in particular the institutions participating in the Gaia Multilateral Agreement.

\section*{Data Availability}

The code and data generated to produce the figures will be shared on reasonable request to NB.



\bibliographystyle{mnras}
\bibliography{My_Library, robyates} 




\appendix

\section{Table of Group Properties and Histograms}\label{sec:appendix}

\begin{table*}
    \centering
    \begin{tabular}{| c | p{1cm} | p{6cm} | p{3cm} | p{2cm} | p{3cm} |}
        \hline
        Group & Sample Size & Chemistry (dex) ($\mu \pm \sigma$) & Kinematics ($L_{z}$ kpc kms$^{-1}$) ($\mu \pm \sigma$) & PC1 - PC2 Location & Notes \\
        \hline
        G1 & 763 & 
        [Ba/Fe]$ = -0.32 \pm 0.21$, [Mg/Fe]$ = 0.32 \pm 0.11$, [Mn/Fe]$ = -0.13 \pm 0.14$, [Y/Fe]$ = -0.02 \pm 0.23$, [Eu/Fe]$ = 0.30 \pm 0.10$, [Zn/Fe]$ = 0.12 \pm 0.22$, [Al/Fe]$ = 0.37 \pm 0.11$, [Fe/H]$ = -0.62 \pm 0.10$, [Cu/Fe]$ = 0.09 \pm 0.10$, [Ti/Fe]$ = 0.28 \pm 0.10$, [Ni/Fe]$ = 0.07 \pm 0.07$, [Sc/Fe]$ = 0.17 \pm 0.07$ & 
        $L_{\rm{z}} = 1175.89 \pm 450.05$, $e = 0.29 \pm 0.19$ & 
        Core & 
        Thick disc chemistry and kinematics \\
        \hline
        G2 & 671 & 
        [Ba/Fe]$ = 0.41 \pm 0.19$, [Mg/Fe]$ = 0.13 \pm 0.10$, [Mn/Fe]$ = -0.37 \pm 0.12$, [Y/Fe]$ = 0.11 \pm 0.17$, [Eu/Fe]$ = 0.44 \pm 0.15$, [Zn/Fe]$ = 0.08 \pm 0.15$, [Al/Fe]$ = -0.12 \pm 0.14$, [Fe/H]$ = -1.00 \pm 0.20$,[Cu/Fe]$ = -0.46 \pm 0.12$, [Ti/Fe]$ = 0.14 \pm 0.10$, [Ni/Fe]$ = -0.11 \pm 0.07$, [Sc/Fe]$ = -0.01 \pm 0.07$ & 
        $L_{\rm{z}} = 145.13 \pm 521.97$, $e = 0.80 \pm 0.21$ & 
        branch & 
        GES  \\
        \hline
        G3 & 651 & 
        [Ba/Fe]$ = 0.44 \pm 0.24$, [Mg/Fe]$ = 0.32 \pm 0.08$, [Mn/Fe]$ = -0.27 \pm 0.10$, [Y/Fe]$ = 0.33 \pm 0.17$, [Eu/Fe]$ = 0.30 \pm 0.10$, [Zn/Fe]$ = 0.19 \pm 0.15$, [Al/Fe]$ = 0.25 \pm 0.15$, [Fe/H]$ = -0.87 \pm 0.18$,[Cu/Fe]$ = -0.14 \pm 0.14$, [Ti/Fe]$ = 0.28 \pm 0.08$, [Ni/Fe]$ = 0.00 \pm 0.08$, [Sc/Fe]$ = 0.08 \pm 0.08$ & 
        $L_{\rm{z}} = 660.54 \pm 599.88$, $e = 0.49 \pm 0.26$ & 
        Core/branch & 
        Bi-modal $L_{\rm{z}}$, in-situ population\\
        \hline
        G4 & 648 & 
        [Ba/Fe]$ = 0.28 \pm 0.12$, [Mg/Fe]$ = 0.31 \pm 0.06$, [Mn/Fe]$ = -0.17 \pm 0.06$, [Y/Fe]$ = 0.12 \pm 0.13$, [Eu/Fe]$ = 0.24 \pm 0.09$, [Zn/Fe]$ = 0.42 \pm 0.08$, [Al/Fe]$ = 0.22 \pm 0.08$, [Fe/H]$ = -0.58 \pm 0.07$,[Cu/Fe]$ = -0.08 \pm 0.07$, [Ti/Fe]$ = 0.20 \pm 0.06$, [Ni/Fe]$ = 0.00 \pm 0.06$, [Sc/Fe]$ = 0.13 \pm 0.04$ & 
        $L_{\rm{z}} = 1218.60 \pm 441.24$, $e = 0.32 \pm 0.20$ & 
        Core & 
        Thick disc kinematics, high [Zn/Fe] distinguished from other thick disc groups \\
        \hline
        G5 & 141 & 
        [Ba/Fe]$ = 0.97 \pm 0.37$, [Mg/Fe]$ = 0.34 \pm 0.12$, [Mn/Fe]$ = -0.21 \pm 0.11$, [Y/Fe]$ = 0.68 \pm 0.26$, [Eu/Fe]$ = 0.39 \pm 0.13$, [Zn/Fe]$ = 0.25 \pm 0.17$, [Al/Fe]$ = 0.31 \pm 0.14$, [Fe/H]$ = -0.75 \pm 0.18$,[Cu/Fe]$ = -0.04 \pm 0.15$, [Ti/Fe]$ = 0.28 \pm 0.08$, [Ni/Fe]$ = 0.02 \pm 0.08$, [Sc/Fe]$ = 0.11 \pm 0.06$ & 
        $L_{\rm{z}} = 840.21 \pm 609.00$, $e = 0.45 \pm 0.25$ & 
        Core/spray & 
        High s-process enhancement \\
        \hline
        G6 & 2235 & 
        [Ba/Fe]$ = -0.05 \pm 0.12$, [Mg/Fe]$ = 0.29 \pm 0.07$, [Mn/Fe]$ = -0.14 \pm 0.06$, [Y/Fe]$ = -0.06 \pm 0.13$, [Eu/Fe]$ = 0.31 \pm 0.09$, [Zn/Fe]$ = 0.13 \pm 0.08$, [Al/Fe]$ = 0.37 \pm 0.06$, [Fe/H]$ = -0.62 \pm 0.08$, [Cu/Fe]$ = 0.08 \pm 0.06$, [Ti/Fe]$ = 0.27 \pm 0.06$, [Ni/Fe]$ = 0.07 \pm 0.05$, [Sc/Fe]$ = 0.14 \pm 0.04$ & 
        $L_{\rm{z}} = 1245.04 \pm 430.14$, $e = 0.31 \pm 0.19$ & 
        Core & 
        Thick disc, chemically similar to G1 and G8 \\
        \hline
        G7 & 832 & 
        [Ba/Fe]$ = -0.10 \pm 0.13$, [Mg/Fe]$ = 0.27 \pm 0.08$, [Mn/Fe]$ = -0.23 \pm 0.07$, [Y/Fe]$ = 0.09 \pm 0.20$, [Eu/Fe]$ = 0.34 \pm 0.06$, [Zn/Fe]$ = -0.08 \pm 0.19$, [Al/Fe]$ = 0.43 \pm 0.06$, [Fe/H]$ = -0.63 \pm 0.09$, [Cu/Fe]$ = 0.12 \pm 0.06$, [Ti/Fe]$ = 0.24 \pm 0.06$, [Ni/Fe]$ = 0.06 \pm 0.05$, [Sc/Fe]$ = 0.18 \pm 0.05$ & 
        $L_{\rm{z}} = 1153.59 \pm 416.14$, $e = 0.31 \pm 0.18$ &
        Core & 
        Thick disc kinematics with bi-modal [Zn/Fe] \\
        \hline
        G8 & 610 & 
        [Ba/Fe]$ = 0.21 \pm 0.15$, [Mg/Fe]$ = 0.36 \pm 0.08$, [Mn/Fe]$ = -0.18 \pm 0.08$, [Y/Fe]$ = -0.05 \pm 0.18$, [Eu/Fe]$ = 0.22 \pm 0.07$, [Zn/Fe]$ = 0.15 \pm 0.15$, [Al/Fe]$ = 0.27 \pm 0.07$, [Fe/H]$ = -0.61 \pm 0.08$, [Cu/Fe]$ = 0.06 \pm 0.08$, [Ti/Fe]$ = 0.19 \pm 0.06$, [Ni/Fe]$ = 0.10 \pm 0.05$, [Sc/Fe]$ = 0.04 \pm 0.05$ & 
        $L_{\rm{z}} = 1073.85 \pm 462.95$, $e = 0.35 \pm 0.21$ &
        Core & 
        Thick disc, chemically similar to G1 and G6 \\
        \hline
        G9 & 581 & 
        [Ba/Fe]$ = 0.24 \pm 0.19$, [Mg/Fe]$ = 0.10 \pm 0.07$, [Mn/Fe]$ = -0.08 \pm 0.08$, [Y/Fe]$ = -0.06 \pm 0.18$, [Eu/Fe]$ = 0.15 \pm 0.09$, [Zn/Fe]$ = -0.03 \pm 0.20$, [Al/Fe]$ = 0.21 \pm 0.09$, [Fe/H]$ = -0.56 \pm 0.06$, [Cu/Fe]$ = 0.05 \pm 0.09$, [Ti/Fe]$ = 0.09 \pm 0.07$, [Ni/Fe]$ = 0.04 \pm 0.07$, [Sc/Fe]$ = 0.06 \pm 0.06$ & 
        $L_{\rm{z}} = 2020.10 \pm 649.82$, $e = 0.19 \pm 0.19$ & 
        Core & 
        Most $\alpha$-poor, thin disc \\
        \hline
        G10 & 365 & 
        [Ba/Fe]$ = 0.63 \pm 0.65$, [Mg/Fe]$ = 0.28 \pm 0.23$, [Mn/Fe]$ = -0.19 \pm 0.28$, [Y/Fe]$ = 0.37 \pm 0.53$, [Eu/Fe]$ = 0.39 \pm 0.24$, [Zn/Fe]$ = 0.26 \pm 0.34$, [Al/Fe]$ = 0.24 \pm 0.28$, [Fe/H]$ = -0.78 \pm 0.22$, [Cu/Fe]$ = -0.11 \pm 0.27$, [Ti/Fe]$ = 0.22 \pm 0.21$, [Ni/Fe]$ = 0.00 \pm 0.16$, [Sc/Fe]$ = 0.12 \pm 0.17$ & 
        $L_{\rm{z}} = 982.75 \pm 831.54$, $e = 0.44 \pm 0.31$ & 
        Core/branch/spray & 
        Significant variance, overlaps with multiple groups \\
        \hline
    \end{tabular}
    \caption{Group characteristics based on sample size, chemistry, kinematics, PC1 - PC2 location, and notable information.}
    \label{tab:group_characteristics}
\end{table*}

\begin{figure*}
  \centering
  \begin{subfigure}
    \centering
    \includegraphics[width=\linewidth]{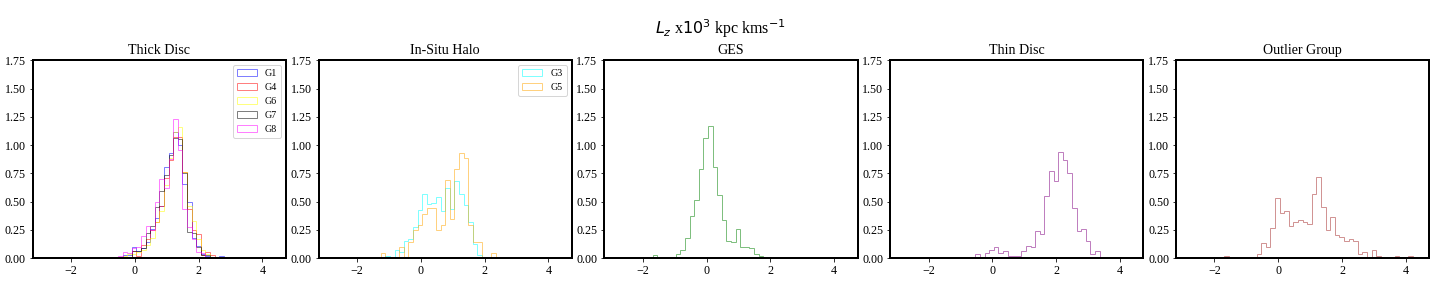}
  \end{subfigure}
  \vspace{-0.5cm} 
  \begin{subfigure}
    \centering
    \includegraphics[width=\linewidth]{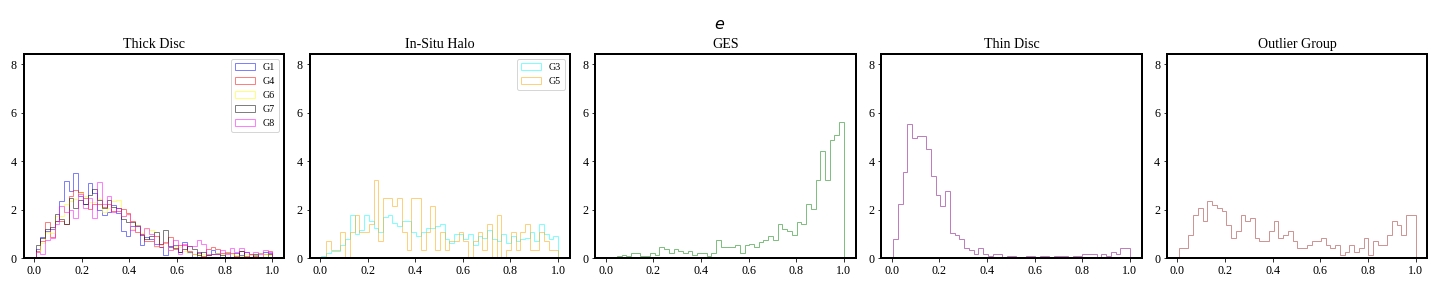}
  \end{subfigure}
   \vspace{-0.5cm} 
  \begin{subfigure}
    \centering
    \includegraphics[width=\linewidth]{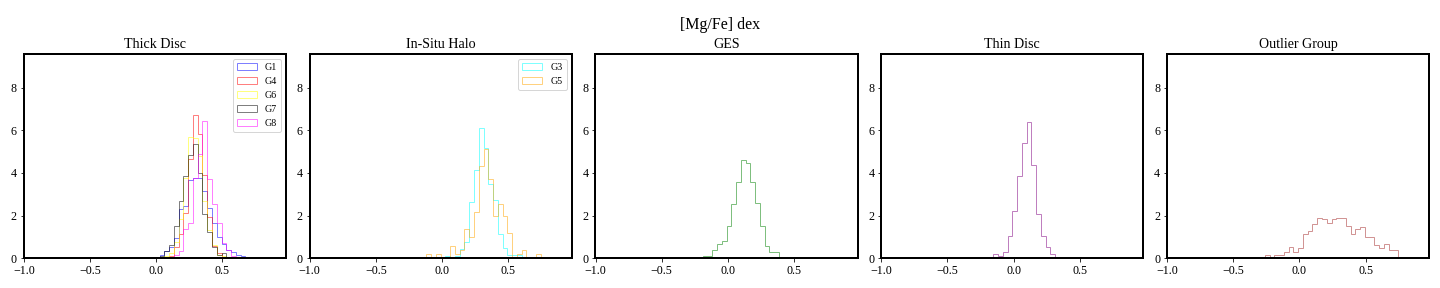}
  \end{subfigure}
  \vspace{-0.5cm} 
  \begin{subfigure}
    \centering
    \includegraphics[width=\linewidth]{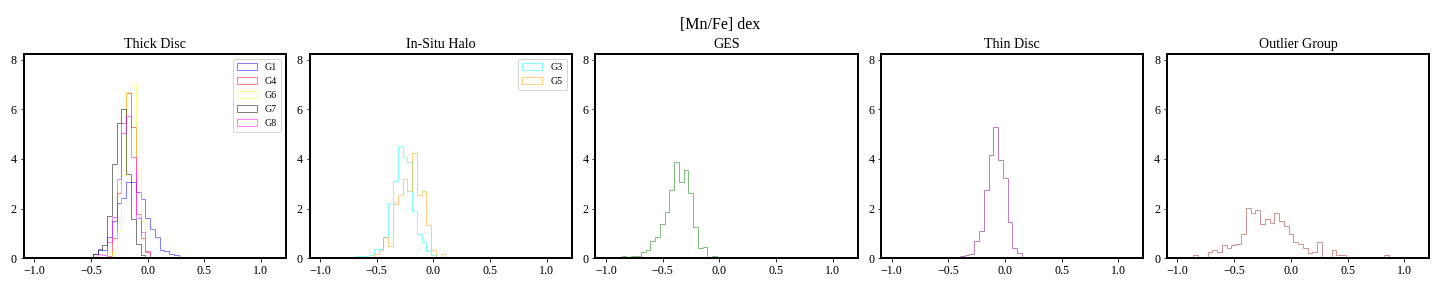}
  \end{subfigure}
  \vspace{-0.5cm} 
  \begin{subfigure}
    \centering
    \includegraphics[width=\linewidth]{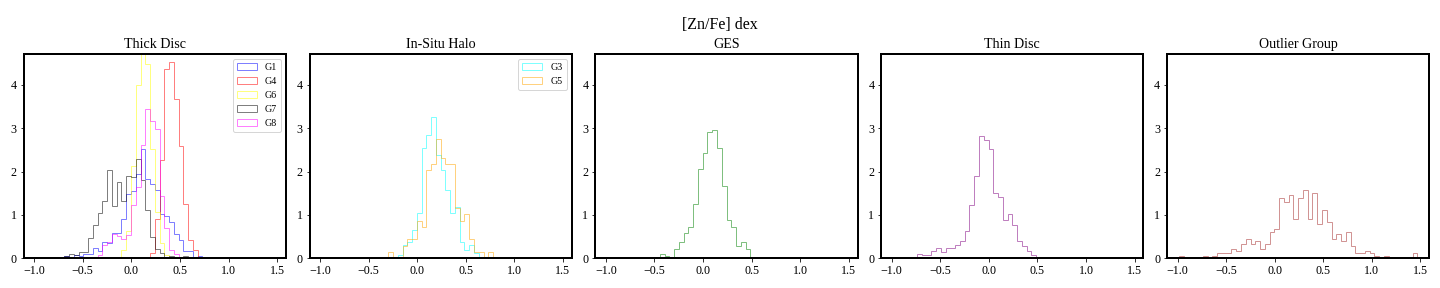}
  \end{subfigure}
  \vspace{-0.5cm} 
  \begin{subfigure}
    \centering
    \includegraphics[width=\linewidth]{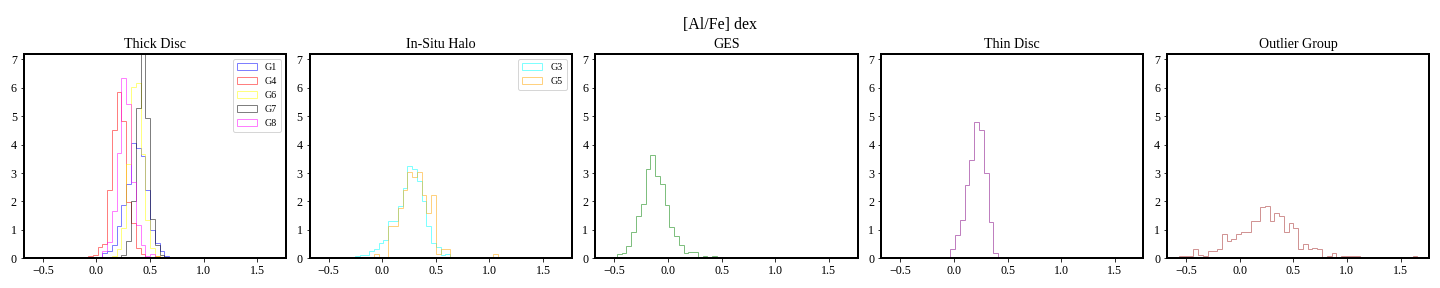}
  \end{subfigure}
  \vspace{-0.5cm} 
  \begin{subfigure}
    \centering
    \includegraphics[width=\linewidth]{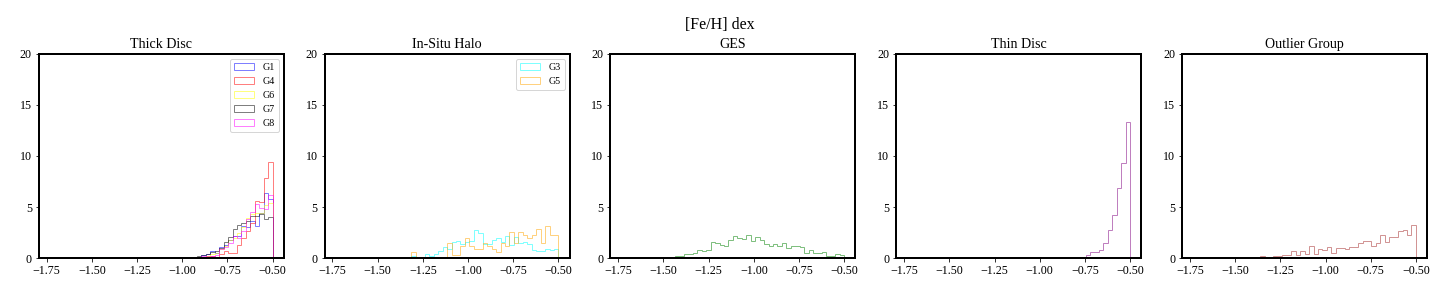}
  \end{subfigure}
\end{figure*}

\begin{figure*}
  \centering
  \vspace{-0.5cm} 
  \begin{subfigure}
    \centering
    \includegraphics[width=\linewidth]{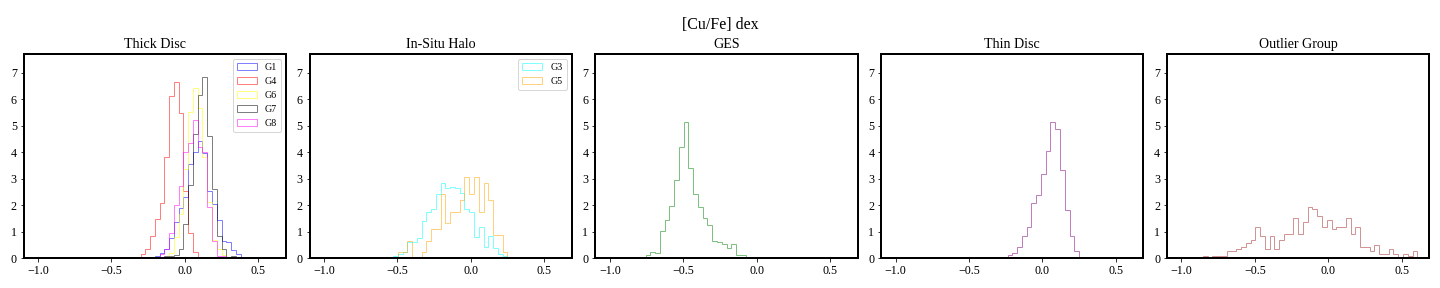}
  \end{subfigure}
  \vspace{-0.5cm} 
  \begin{subfigure}
    \centering
    \includegraphics[width=\linewidth]{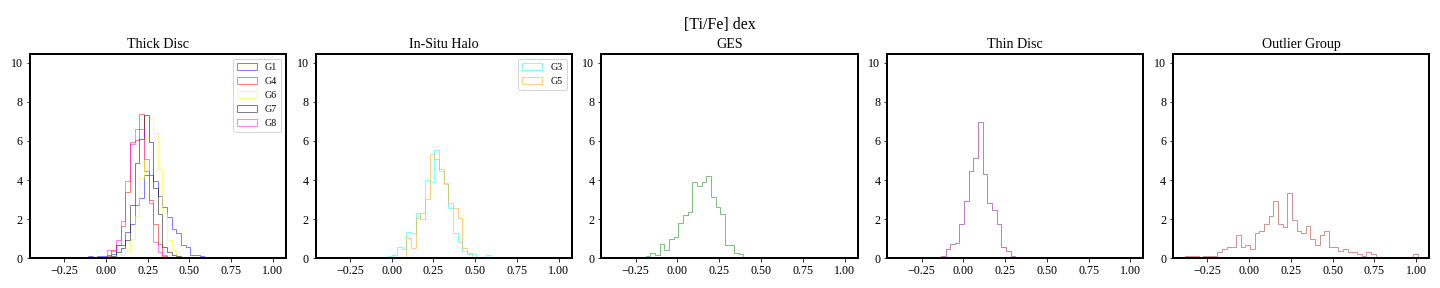}
  \end{subfigure}
  \vspace{-0.5cm} 
  \begin{subfigure}
    \centering
    \includegraphics[width=\linewidth]{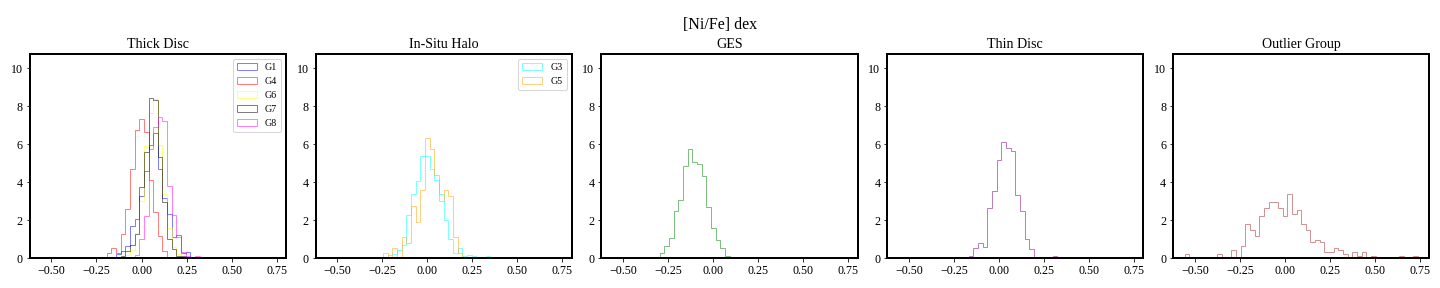}
  \end{subfigure}
  \vspace{-0.5cm} 
  \begin{subfigure}
    \centering
    \includegraphics[width=\linewidth]{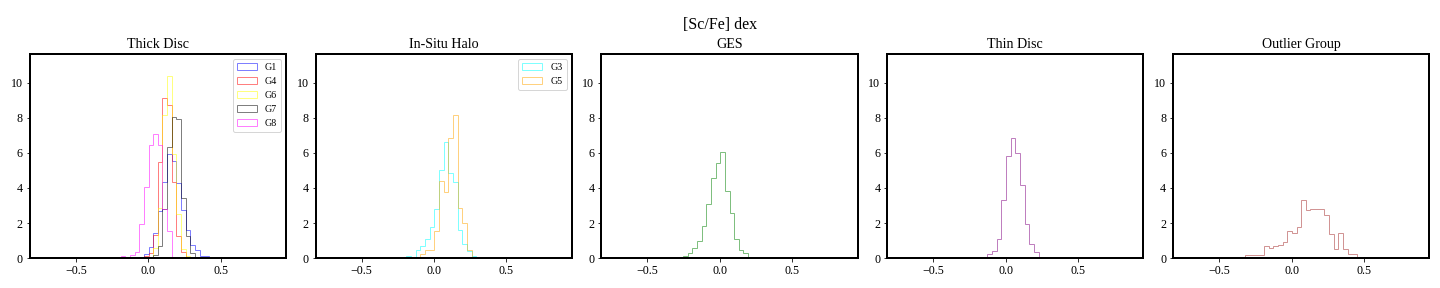}
  \end{subfigure}
\vspace{-0.5cm} 
  \begin{subfigure}
    \centering
    \includegraphics[width=\linewidth]{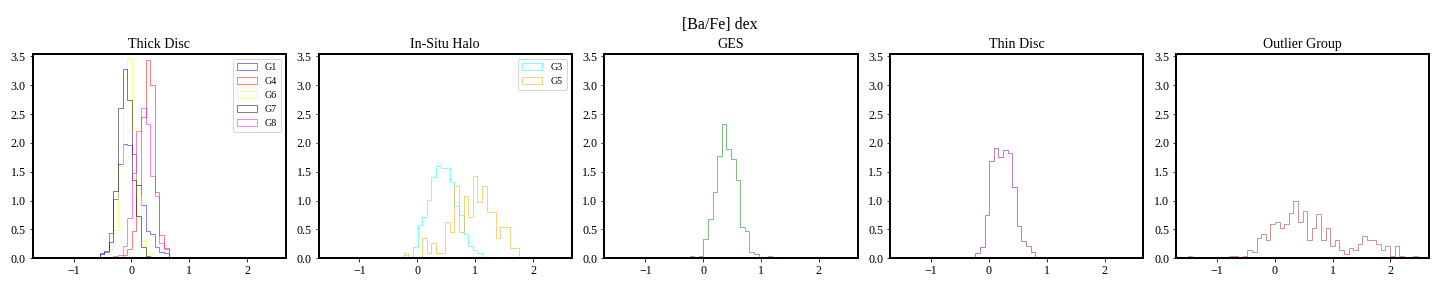}
  \end{subfigure}
  \vspace{-0.5cm} 
  \begin{subfigure}
    \centering
    \includegraphics[width=\linewidth]{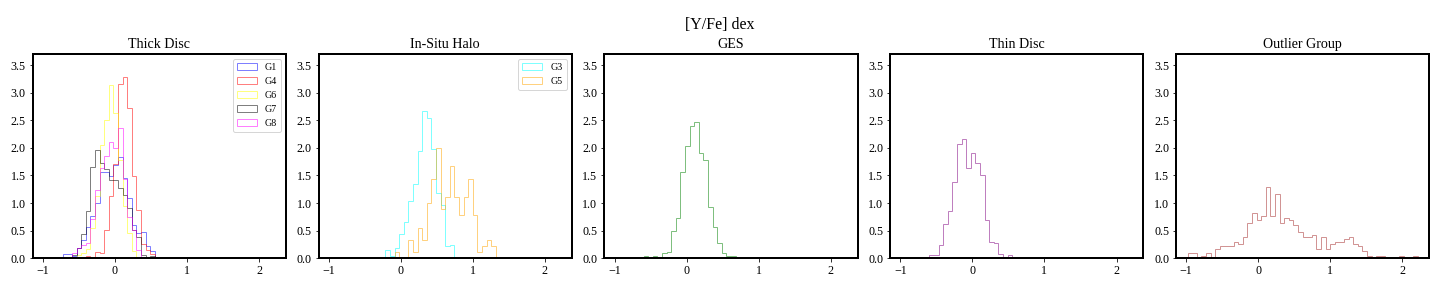}
  \end{subfigure}
  \vspace{-0.5cm} 
  \begin{subfigure}
    \centering
    \includegraphics[width=\linewidth]{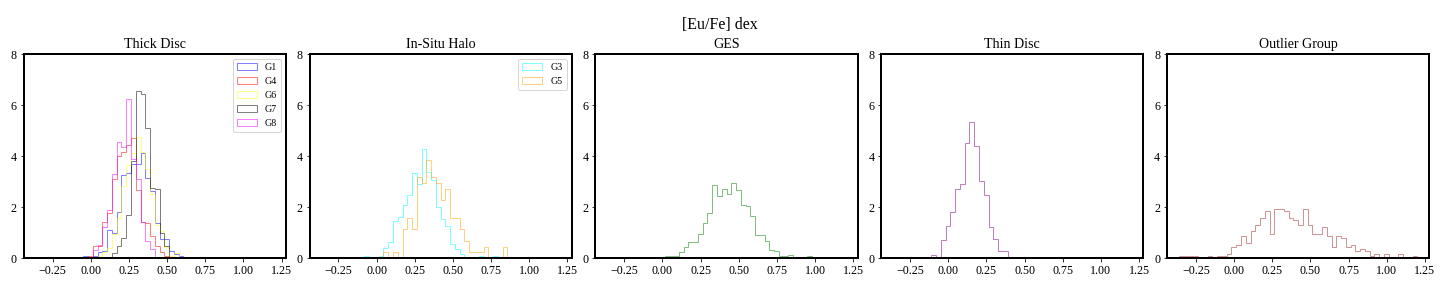}
  \end{subfigure}
\end{figure*}


\bsp	
\label{lastpage}
\end{document}